\documentclass{emulateapj}
\usepackage{graphicx}

\slugcomment{Accepted to MNRAS}

\begin{document}

\title{The Cygnus Loop's Distance, Properties, \& Environment Driven Morphology} 

\author{Robert A.\ Fesen, Kathryn E.\ Weil \&  Ignacio A.\ Cisneros\footnote{Visiting Researcher} }
\affil{6127 Wilder Lab, Department of Physics \& Astronomy, Dartmouth
College, Hanover, NH 03755, USA} 
\author{William P.\ Blair}
\affiliation{Henry A. Rowland Department of Physics \& Astronomy, Johns Hopkins University, 3400 N. Charles Street, Baltimore, MD 21218, USA}
\author{John C.\ Raymond}
\affiliation{Harvard-Smithsonian Center for Astrophysics, 60 Garden St., Cambridge, MA 02138, USA}

\begin{abstract}

The Cygnus Loop is among the brightest and best studied evolved Galactic
supernova remnants. However, its distance has remained uncertain thus
undermining quantitative understanding about many of its fundamental
properties.  Here we present moderate-dispersion spectra of stars with
projected locations toward the remnant.  Spectra  of three stars revealed
\ion{Na}{1} 5890,5896 \AA \ and \ion{Ca}{2} 3934 \AA \ absorption features
associated with the remnant's expanding shell, with velocities ranging from
$-160$ to +240 km s$^{-1}$.  Combining Gaia DR2 parallax measurements for these
stars with other recent observations, we find the distance to the Cygnus Loop's
centre is $735 \pm 25$ pc, only a bit less than the 770 pc value proposed by
Minkowski some 60 years ago. Using this new distance, we discuss the remnant's
physical properties including size, SN explosion energy, and shock velocities.
We also present multi-wavelength emission maps which reveal that, instead of
being located in a progenitor wind-driven cavity as has long been assumed, the
Cygnus Loop lies in an extended, low density region. Rather than wind-driven
cavity walls, these images reveal in unprecedented clarity the sizes and
locations of local interstellar clouds with which the remnant is interacting,
giving rise to its large-scale morphology.

\end{abstract}



\section{Background}

The Galactic supernova remnant (SNR) G74.0-8.5, commonly known as the Cygnus
Loop or Veil Nebula, is thought to be a middle-age remnant with an estimated
age of $1 - 2 \times 10^{4}$ yr.  Based on extensive multi-wavelength
observations, the Cygnus Loop has been most often modeled as the remnant of a
supernova explosion that occurred inside an interstellar cavity likely created
by a high-mass progenitor star
\citep{McCray1979,Hester1994,Levenson1997,Levenson1998,MT1999,Fang2017}.

\begin{figure*}
\begin{center}
\includegraphics[width=1.0\textwidth]{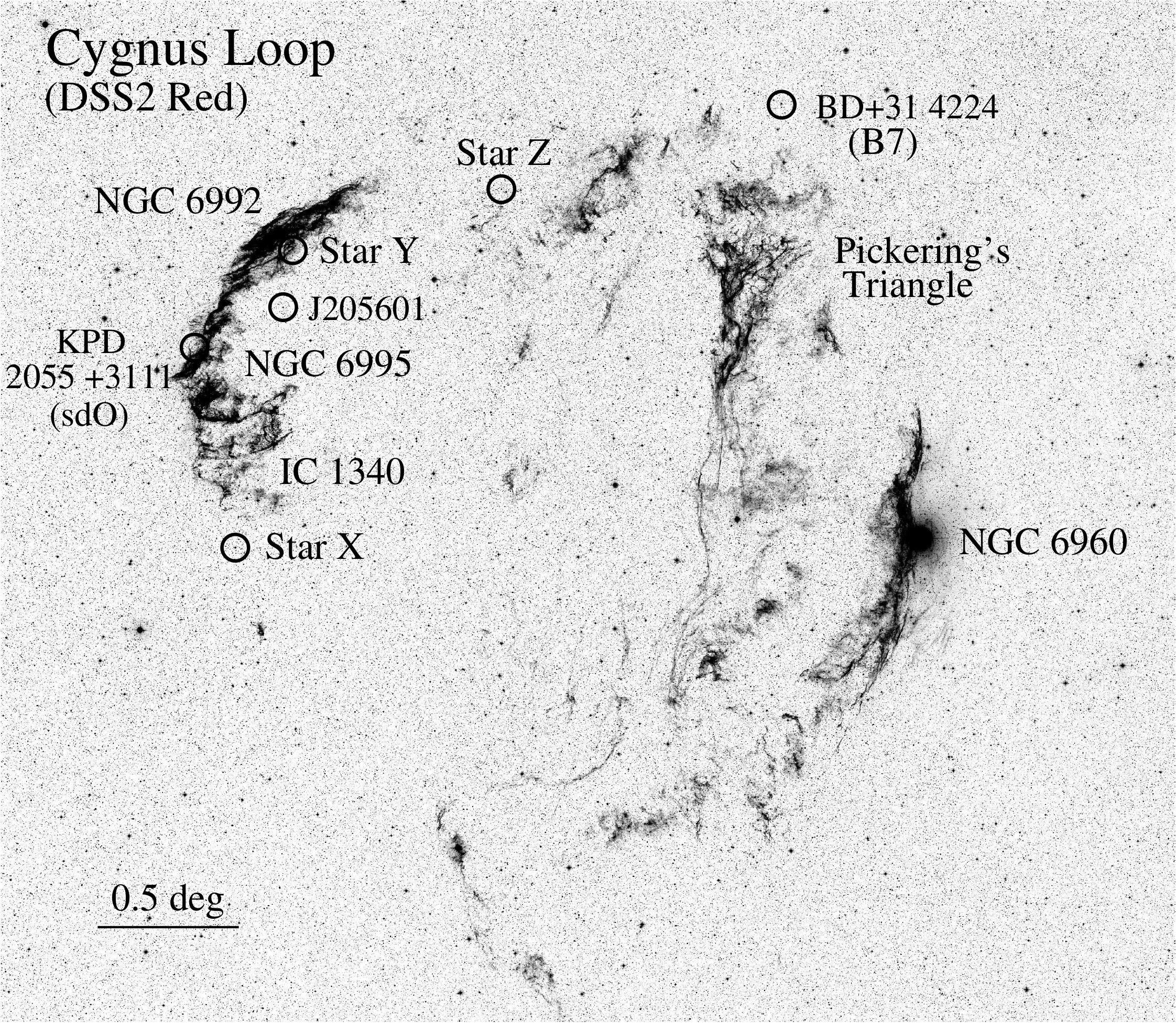}
\caption{Reproduction of the red image of the Digital Sky Survey of the Cygnus
Loop.  Marked are our three program stars X, Y, and Z, along with the
\citet{Blair2009} sdO star (KPD 2055 +3111), and the two stars, J205601
\& BD+31 4224, suspected by \citet{Fesen2018} to be physically interacting with
the remnant's shock front.  North is up, East to the left.  } 
\end{center}
\label{DSS2} 
\end{figure*}

Underlying any quantitative analysis of the Cygnus Loop is uncertainty about
its distance.  For many decades, the most often adopted value 
was a kinematic investigation by \citet{Minkowski1958}.  \citet{Hubble1937}
measured a proper motion of $0\farcs03$ yr$^{-1}$ for the remnant's bright
eastern and western nebulae (see Fig.\ 1).  This value, when combined with an average radial
velocity of 115 km s$^{-1}$ for the remnant's optical filaments, led
\citet{Minkowski1958} to estimate a distance of $770$ pc.

\begin{deluxetable*}{llcccccr}
\tablecolumns{8}
\tablecaption{Stars Used to Estimate the Distance to the Cygnus Loop}
\tablewidth{0pt}
\tablehead{ \colhead{Star} & \colhead{Catalog} & \colhead{RA}      & \colhead{Dec}     & \colhead{Magnitude} & \colhead{Gaia DR2} & \colhead{Gaia DR2} & \colhead{Location}  \\
            \colhead{ID\tablenotemark{a}}   & \colhead{Number\tablenotemark{b}}    & \colhead{(J2000)} & \colhead{(J2000)} & \colhead{(v)}       & \colhead{Parallax}   &   \colhead{Distance} &  \colhead{to SNR}  }
\startdata
 J205601         & TYC 2688-1037-1  & 20:56:00.936  & +31:31:29.75  & 11.57  & $0.6334\pm 0.0661$ mas &$1580 \pm180$ pc      & far behind  \\
 KPD 2055 +3111  & UCAC2 42838052   & 20:57:26.889  & +31:22:52.56  & 14.12  & $1.2610\pm 0.0413$ mas & $793 \pm 30$ pc      &      behind \\
 BD+31 4224      & TYC 2691-290-1   & 20:47:51.817  & +32:14:11.33  &~~9.58  & $1.3033\pm 0.0438$ mas & $767 \pm 27$ pc      &     inside \\
 Star X          & HD 335334        & 20:56:44.629  & +30:41:14.33  &~~9.51  & $1.3586\pm 0.0440$ mas & $736\pm25$ pc        & just behind  \\
 Star Y          & TYC 2688-365-1   & 20:55:51.948  & +31:43:27.40  & 11.25  & $1.3598\pm 0.0449$ mas & $735\pm25$ pc        & just behind  \\
 Star Z          & TYC 2692-3378-1  & 20:52:27.557  & +31:56:29.48  & 10.73  & $1.1581\pm 0.0390$ mas & $864\pm30$ pc        & far behind   \\
\enddata
\tablenotetext{a}{References: J205601 \& BD+31 4224: \citet{Fesen2018}; KPD 2055 +3111: \citet{Blair2009}; X, Y, Z: this work}
\tablenotetext{b}{TYC: \citet{Hog2000}; UCACA2: \citet{Zach2004}; HD: Henry Draper Catalogue }
\end{deluxetable*}

However more recent distance estimates have ranged from 440 to 1400 pc (see
review by \citealt{Fesen2018}).  The most cited value is $540^{+100}_{-80}$ pc
by \citet{Blair2005}. Although smaller than many other estimates, this
distance appeared in line with a subsequent study which estimated a distance of $576 \pm 61$
pc to a sdO star, KPD 2055 +3111, lying in the direction of the remnant's
eastern nebulosities \citep{Blair2009}. 

Because this star's UV spectrum showed high-velocity O VI 1032 \AA \ line
absorption,  \citet{Blair2009} argued the star must be behind the remnant thus
limiting the remnant's distance to less than $\simeq$ 576 pc. However, Gaia DR2 data
(\citealt{Gaia2018}) report a parallax of $1.2610 \pm 0.0413$ mas and  hence a
distance of $793 \pm 30$ pc for this star (Table 1).

Other recent distance estimates have tended to favor larger values.
\citet{Medina2014} used spectra of Balmer-dominated H$\alpha$ filaments to
estimate shock velocities of around 400 km s$^{-1}$ based on H$\alpha$ emission
profiles.  Combined with proper motions by \citet{Salvesen2009}, they deduced a
distance of around 890 pc. However, a re-analysis of thermal equilibrium in a
collisionless shock led to a reduction of the derived shock velocities which
decreased the remnant's estimated distance to $\simeq$ 800 pc
\citep{Raymond2015}.

The most recent distance estimate to the Cygnus Loop is that of
\citet{Fesen2018} who suggested optical nebulosities around two stars were
stellar mass-loss material interacting with the remnant's expanding shock front
and thus could be used to estimate the Cygnus Loop's distance.  A small
emission-line nebula with a bow-shaped morphology centred on a M4 III star,
J205601, near the remnant's eastern nebula NGC~6992 was shown to be
shock-heated with significantly higher electron densities than typically seen
in the remnant's filaments.  Since distance estimates to this red giant ranged
from 1.0 to 4.2 kpc, \citet{Fesen2018} struggled to connect this star and the
nebulosity to the remnant.  Gaia's DR2 parallax of $0.6334 \pm 0.0661$ mas
implies a distance around 1.6 kpc kpc, placing it well behind the remnant
(Table 1), leaving the origin of the bow-shaped nebulosity somewhat of a
mystery.  

\citet{Fesen2018} also identified a B7 V star, BD+31 4224, located along the remnant's
northwestern limb within a small arc of nebulosity in a complex region of
curved and distorted filaments suggestive of a disturbance of the remnant's
shock front due to the B star's stellar winds. Whereas \citet{Fesen2018}
estimated the star's distance to be 0.8 to 1.0 kpc, Gaia DR2 reports a parallax of
$1.3033 \pm 0.0438$ mas implying a distance of $767 \pm35$ pc (Table 1).

This distance estimate, when taken together with the \citet{Blair2009} sdO star
lying behind the remnant, suggests a firm distance limit of less than 800 pc, 
consistent with Minkowski's 1958 value of 770 pc. A distance $\simeq$ 750 - 800 pc 
could explain the failure by \citet{Welsh2002} to detect high-velocity
\ion{Na}{1} and \ion{Ca}{2} line absorptions associated with the remnant in
several stars located along the line-of-sight to the Cygnus Loop at distances less
than $\sim$700 pc.

In light of the recent Gaia data release giving relatively accurate
distances to many earlier type stars located toward the Cygnus Loop, we 
obtained optical spectra of several stars located between 700 and 900 pc in the
remnant's direction to search for associated high-velocity \ion{Na}{1} and
\ion{Ca}{2} absorption features as additional clues to the remnant's distance.  Our
spectral observations are described in $\S$2, with results presented in $\S$3.
Combining these new data with prior observations, we discuss in $\S$4 the
Cygnus Loop's likely distance and thus its true size and supernova energy.
We also present multi-wavelength images which reveal the remnant's likely interactions
with several local interstellar clouds leading to a re-assessment of the long assumed 
progenitor wind-driven cavity model. Our conclusions are summarized in $\S$5. 

\section{Observations}

Roughly two dozen stars were selected as spectroscopic program targets on the
basis of Gaia determined distances between 700 and 950 pc, m$_{\rm g} <$ 13.5,
Gaia T$_{\rm eff}$ values greater than 8000 K, and projected locations within
the Cygnus Loop's optical boundaries.  Many had early-type A classifications
listed on SIMBAD \citep{Wenger2000}.  This modest target list was viewed as
sufficiently large to uncover at least a few background stars exhibiting
high-velocity interstellar absorption components associated with the Cygnus
Loop.  

Moderate-dispersion spectra of several of these stars were observed on 2018 May 30,
31 and June 1, 2, and 3  with the 1.3m McGraw-Hill telescope at MDM Observatory
located at Kitt Peak, Arizona.  The data were taken using a Boller \& Chivens
Spectrograph (CCDS) which employs a Loral 1200 $\times$ 800 CCD detector. We
used an 1800 grooves mm$^{-1}$ grating blazed at 4700 \AA \ to yield a wavelength
coverage of 330 \AA \ with a spectral scale of 0.275 \AA \ per pixel.  Exposure
times varied from $1-2 \times 1200$ s to $2 \times 3000$ s depending on star
brightness and wavelength coverage.

A 1.5 arcsec wide slit was used which resulted in a measured FWHM of 2.2 pixels,
corresponding to a spectral resolution of 0.61 \AA, providing an effective R
$\simeq$ 9700 at the \ion{Na}{1} 5890,5996 \AA \ lines and R $\simeq$ 6500 at the
\ion{Ca}{2} K 3934 \AA \ line.  Although these spectral resolutions are low
relative to conventional interstellar absorption studies where R values
typically exceed 30,000, this spectrograph + grating system provided
velocity resolutions $\approx$ 30 and 45 km s$^{-1}$ for the \ion{Na}{1} and
\ion{Ca}{2} lines, respectively. This was judged adequate for detecting the expected
high-velocity expansion filaments, in the range of 50 to 300 km s$^{-1}$, in the
spectrum of background stars.

Standard pipeline data reduction was performed using IRAF\footnote{IRAF is
distributed by the National Optical Astronomy Observatories, which is operated
by the Association of Universities for Research in Astronomy, Inc.\ (AURA)
under cooperative agreement with the National Science Foundation.}. The spectra
were bias-subtracted, cosmic-ray corrected using the L.A. Cosmic software
\citep{vanDokkum2001}, co-added, wavelength calibrated using
Hg, Ne, Ar, and Xe comparison lamps, and have been corrected to local standard
of rest (LSR) values. Detected interstellar \ion{Na}{1} and
\ion{Ca}{2} line components were deblended using IRAF Gaussian fitting routines.
Measured velocities have a typical
uncertainties of $\pm 5$ km s$^{-1}$.

\section{Results}

\begin{figure*}[t]
\centering
\includegraphics[height=0.26\textheight]{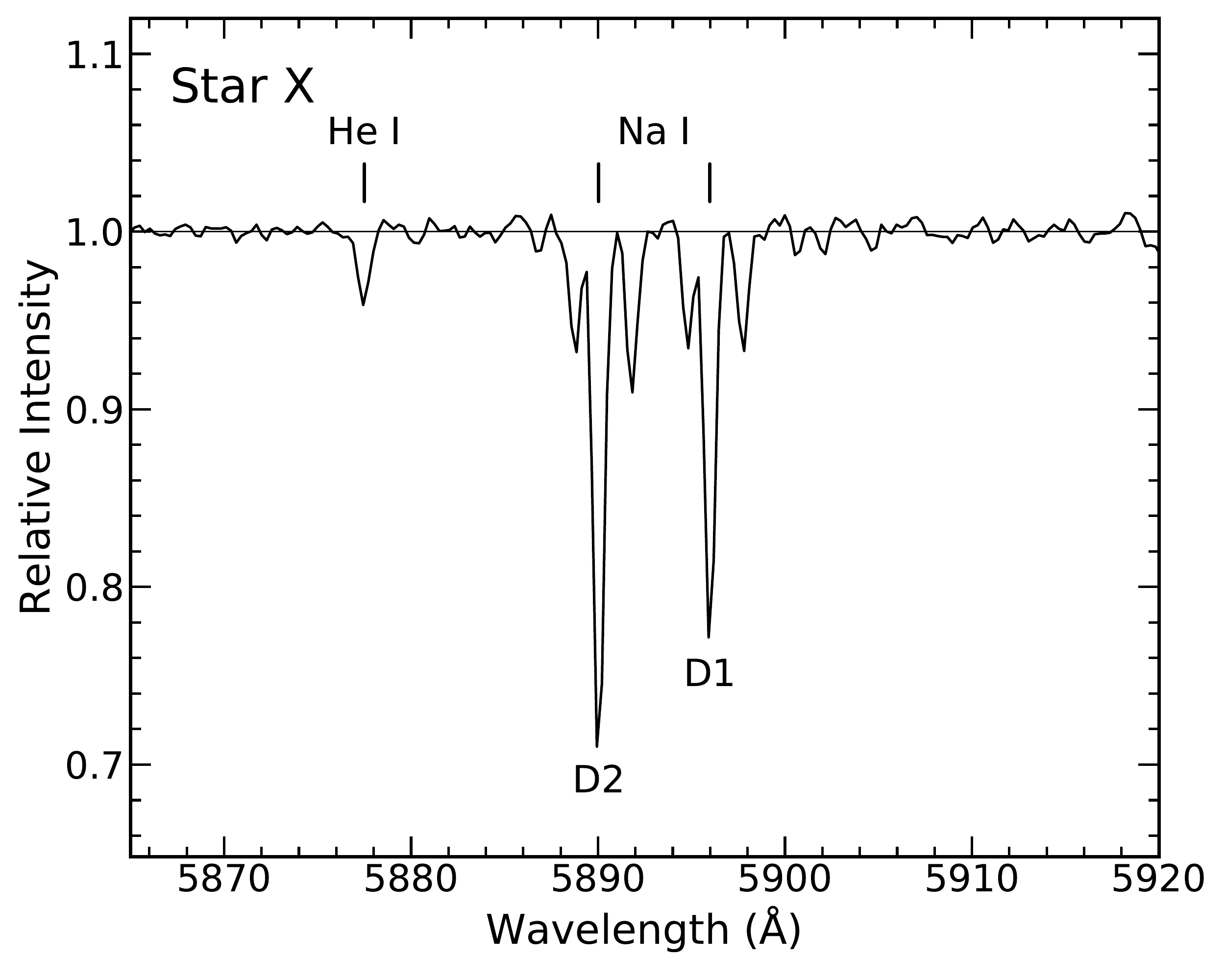}
\includegraphics[height=0.26\textheight]{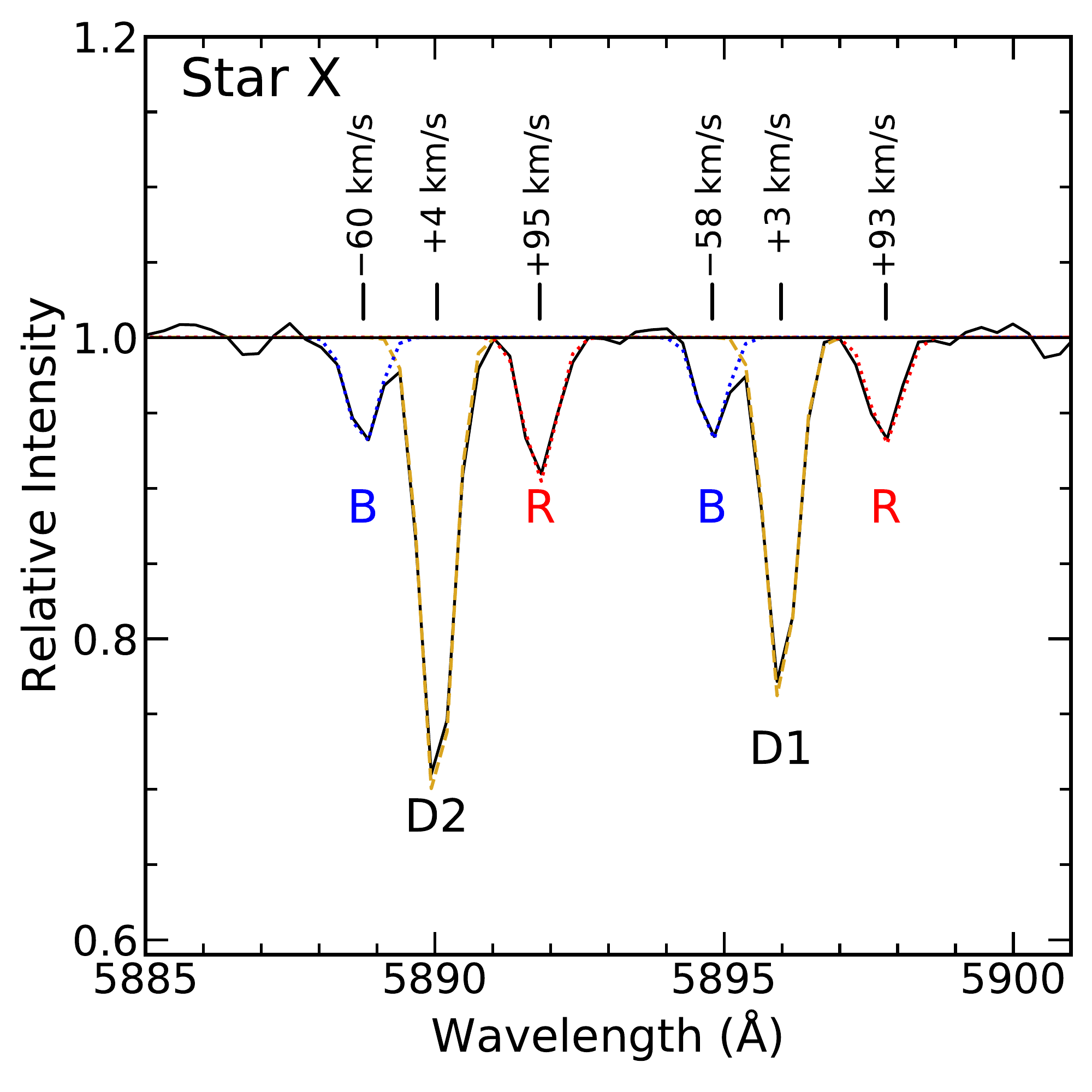}\\
\includegraphics[height=0.26\textheight]{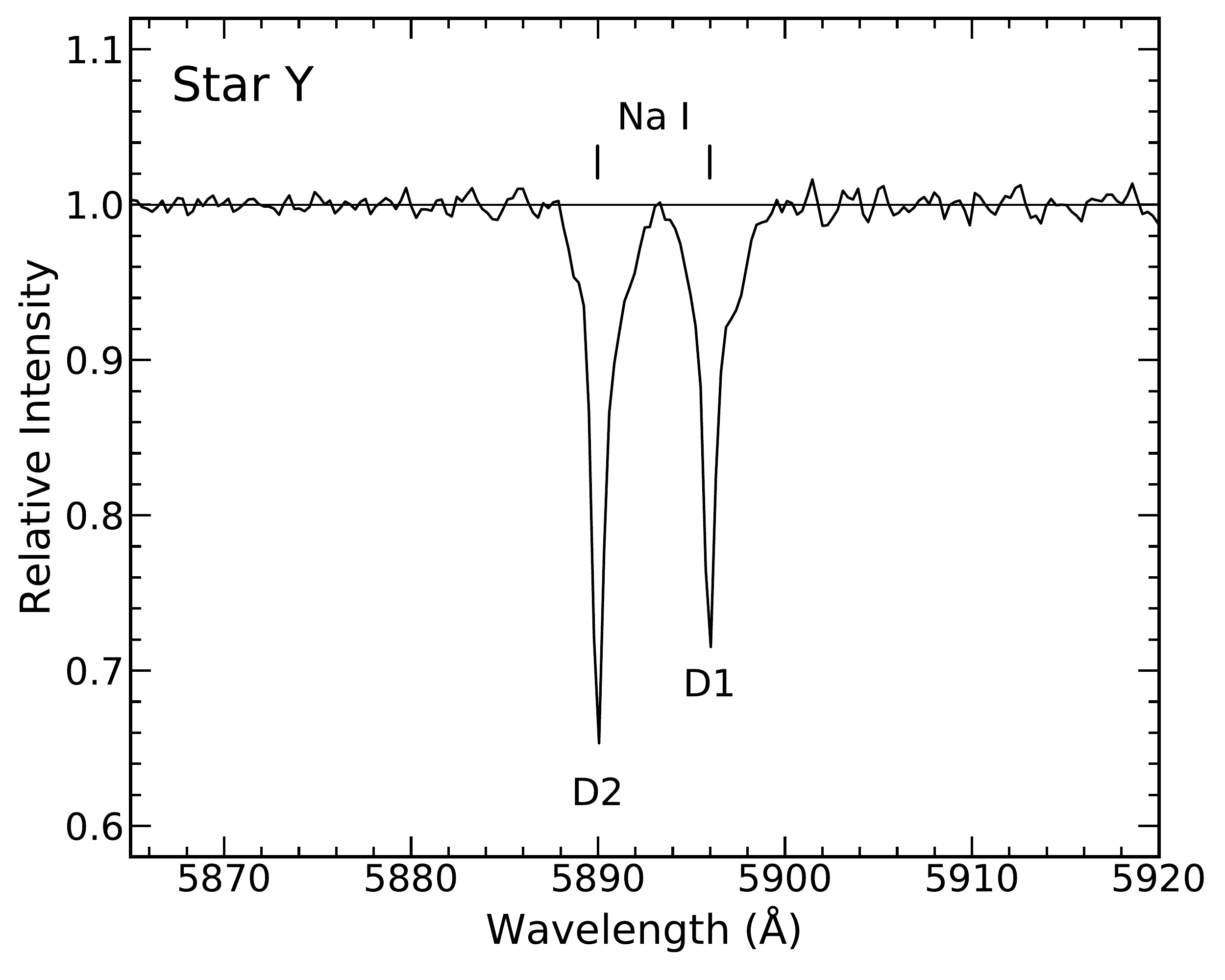}
\includegraphics[height=0.26\textheight]{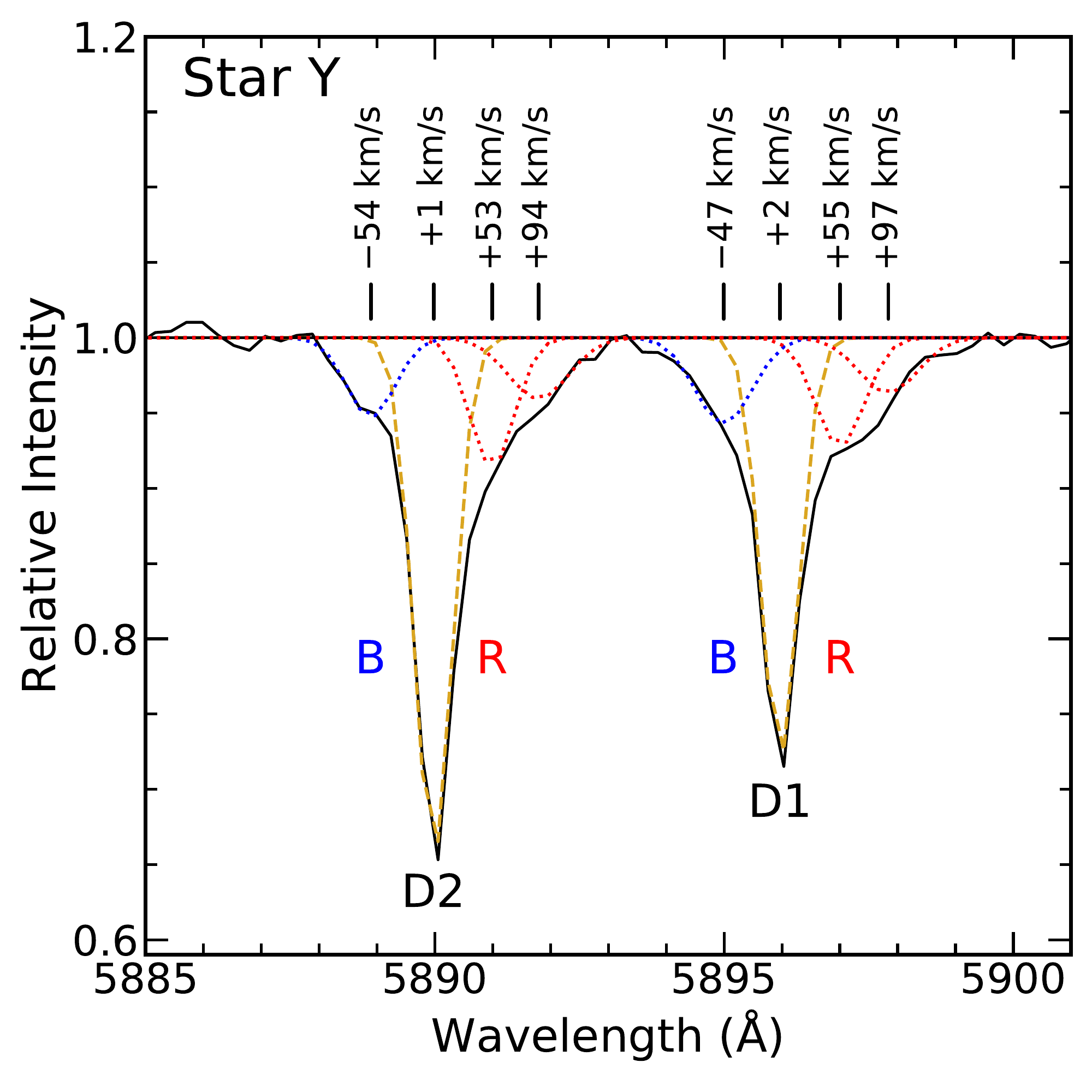}\\
\includegraphics[height=0.26\textheight]{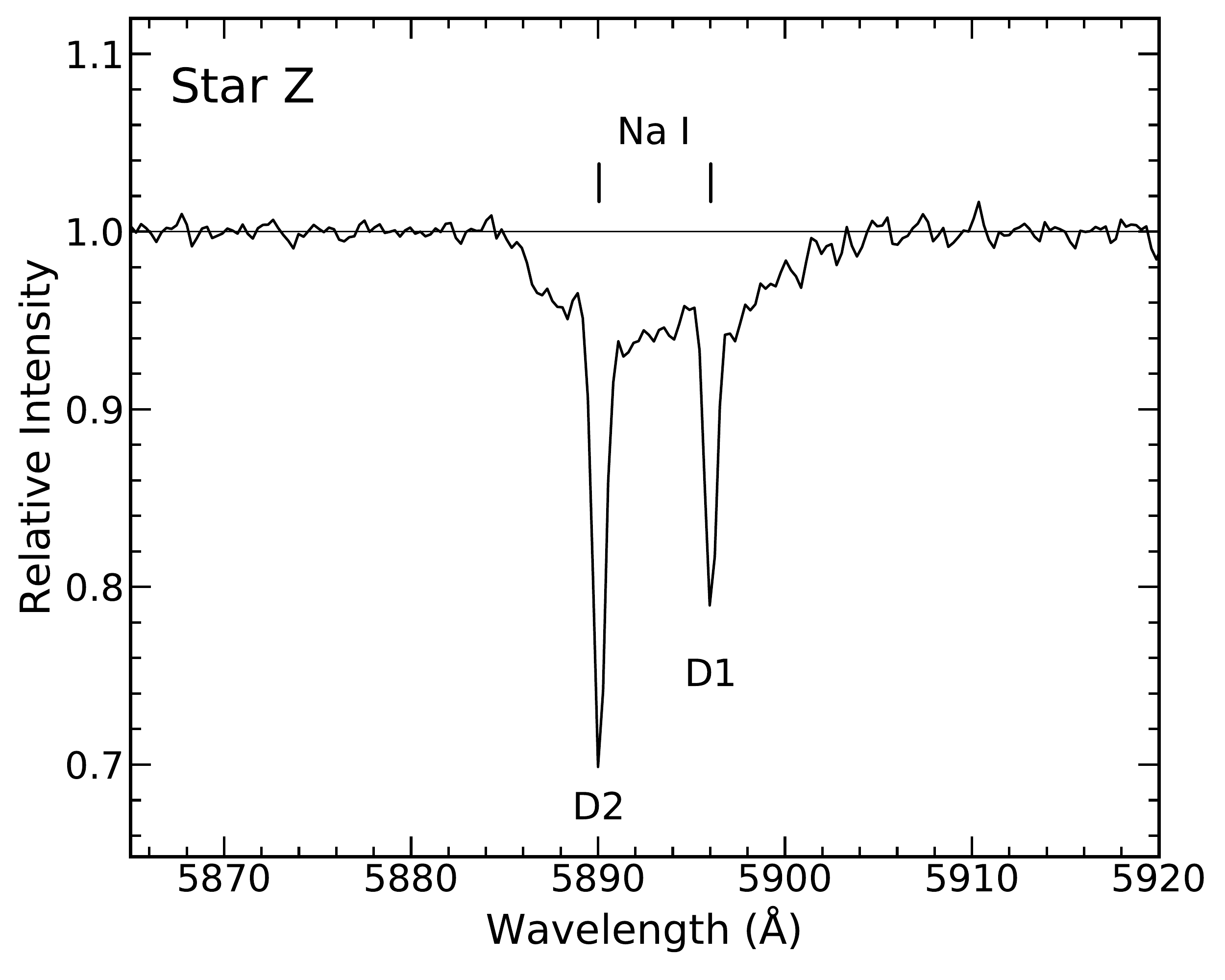}
\includegraphics[height=0.26\textheight]{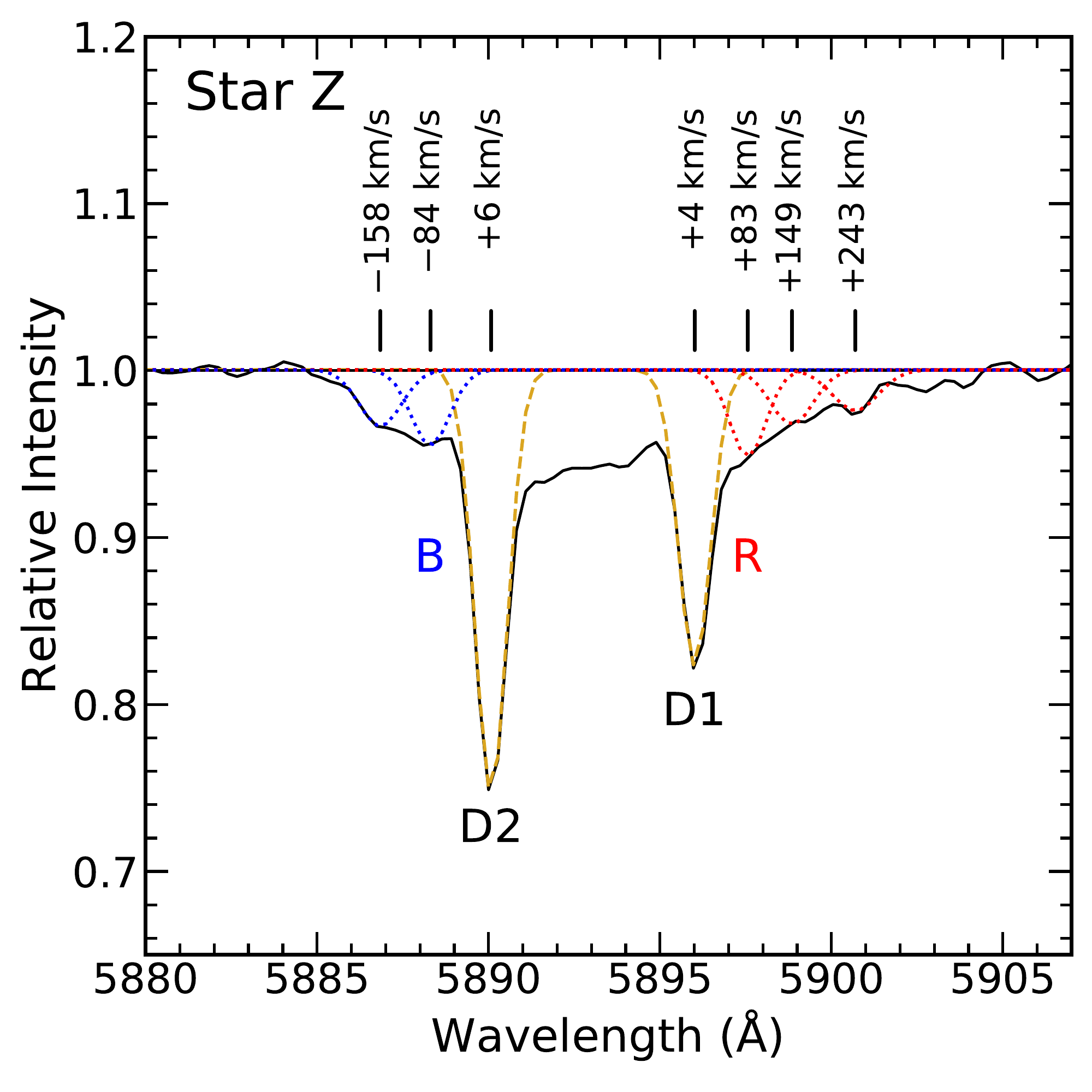}
\caption{Spectra covering the \ion{Na}{1} D line region for Stars X, Y, and Z.
   }
\label{NaI_spectra}
\end{figure*}

\begin{deluxetable*}{llcccccccccccc}[t]
\tablecolumns{18}
\tablecaption{Measured Interstellar Velocity Components and Equivalent Widths}
\tablewidth{0pt}
\tablehead{
 \colhead{Star} & \colhead{Line}  & \colhead{$\mathrm{V}_0$}      & \colhead{$\mathrm{EW}_0$}     & \colhead{$\mathrm{V_{B1}}$} & \colhead{$\mathrm{EW_{B1}}$} & \colhead{$\mathrm{V_{B2}}$} &
 \colhead{$\mathrm{EW_{B2}}$}  & \colhead{$\mathrm{V_{R1}}$} & \colhead{$\mathrm{EW_{R1}}$} & \colhead{$\mathrm{V_{R2}}$} & \colhead{$\mathrm{EW_{R2}}$} & \colhead{$\mathrm{V_{R3}}$} &
 \colhead{$\mathrm{EW_{R3}}$} \\
  \colhead{ID} &  & \colhead{km s$^{-1}$} & \colhead{\AA} &  \colhead{km s$^{-1}$}      & \colhead{\AA}   &    \colhead{km s$^{-1}$} &  \colhead{\AA} &  \colhead{km s$^{-1}$}      & \colhead{\AA}   &    \colhead{km s$^{-1}$} &  \colhead{\AA} &    \colhead{km s$^{-1}$} &  \colhead{\AA}}
\startdata
 Star X &  \ion{Na}{1}  D1      & +3  & 0.17  & $-$58 & 0.04 & \nodata  & \nodata& +93 & 0.05 & \nodata& \nodata  & \nodata &\nodata \\
                &  \ion{Na}{1}  D2      & +4 & 0.22  &  $-$60 & 0.05 & \nodata &\nodata & +95 & 0.07 & \nodata & \nodata & \nodata &\nodata \\
                &  \ion{Ca}{2}  K       & $-$2 & 0.13  &  $-$58 & 0.11 & \nodata & \nodata& +88 & 0.29 &\nodata & \nodata & \nodata & \nodata \\
 Star Y &  \ion{Na}{1}  D1      & +2  & 0.23   &$-$47 & 0.07 & \nodata   & \nodata& +55 & 0.08 & +97 & 0.05  & \nodata & \nodata \\
                &  \ion{Na}{1}  D2      &  +1 & 0.29  &$-$54  & 0.06 & \nodata   & \nodata& +53 & 0.08 & +94 & 0.05 & \nodata & \nodata \\
 Star Z         &  \ion{Na}{1}  D1      & +4  & 0.21 &\nodata &\nodata &\nodata &\nodata & +83 & 0.07 & +149   & 0.05  & +243 & 0.04\\
                &  \ion{Na}{1}  D2      &  +6  & 0.30  & $-$84 & 0.06  &$-$158    & 0.05  & \nodata &\nodata  &\nodata &\nodata    &\nodata &\nodata \\
\enddata
\end{deluxetable*}

Although our program stars were distributed across the remnant, only three
stars located in eastern half were found to exhibit high-velocity absorption
components visible given our spectral resolution and S/N.  Projected locations
of these three stars, labeled X, Y, and Z are shown in Figure~\ref{DSS2} along
with the locations of the sdO star KPD 2055 +3111 (hereafter KPD) studied by
\citet{Blair2009} and the two stars J205601 (M4 III) and BD+31~4224 (B7 V)
discussed by \citet{Fesen2018}. Coordinates and Gaia parallax measurements for
all six stars are listed in Table 1.  While we were unable to classify our
three observed program stars due to our limited spectral coverage, Stars X, Y,
and Z have $B-V$ values of 0.03, 0.08, and 0.29 with Star X classified as A0
\citep{Wenger2000}. 

Reduced and normalized spectra of Stars X, Y, and Z for the wavelength region
around the \ion{Na}{1} D lines at 5889.95 and 5895.92 \AA \  are showed in
Figure~\ref{NaI_spectra}. Measured D1 and D2 absorption component velocities
and equivalent widths are listed in Table 2.  

The spectrum of the 9.5 mag Star X located south of NGC~6995 and IC~1340 exhibited the
cleanest high-velocity absorption components, showing both red and blue
absorption features.  Deconvolution of the observed \ion{Na}{1} profiles
indicates high-velocity interstellar components at $\simeq -60$ and $+95$ km
s$^{-1}$ (Fig.~\ref{NaI_spectra}; upper panels).  With a Gaia estimated distance of $736 \pm
25$ pc, the detection of both red and blue shifted \ion{Na}{1} features in this
star's spectrum implies a distance limit to the backside of the Cygnus Loop of less than
$\simeq$ 760 pc.

\begin{figure}
\centering
\includegraphics[height=0.26\textheight]{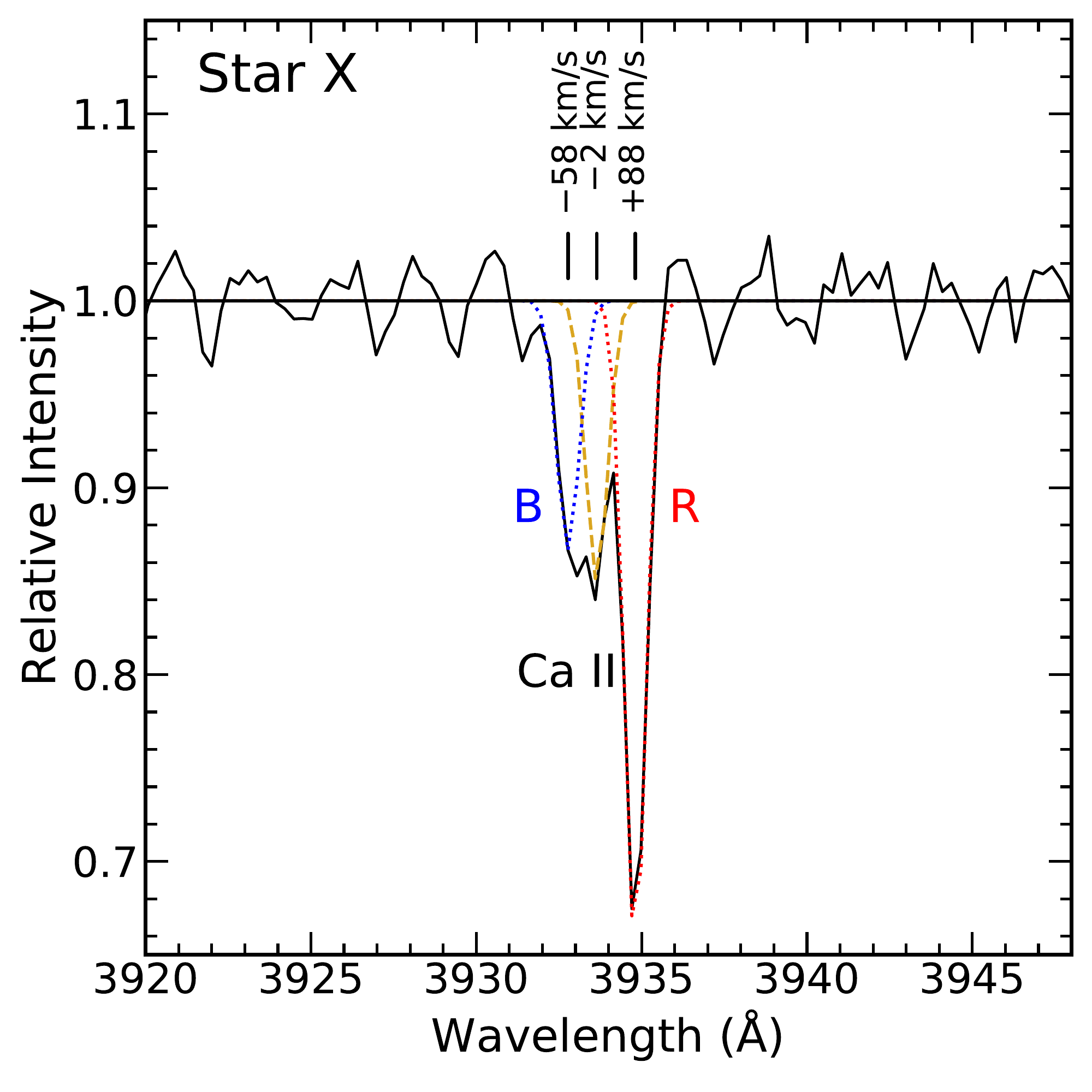}
\includegraphics[height=0.26\textheight]{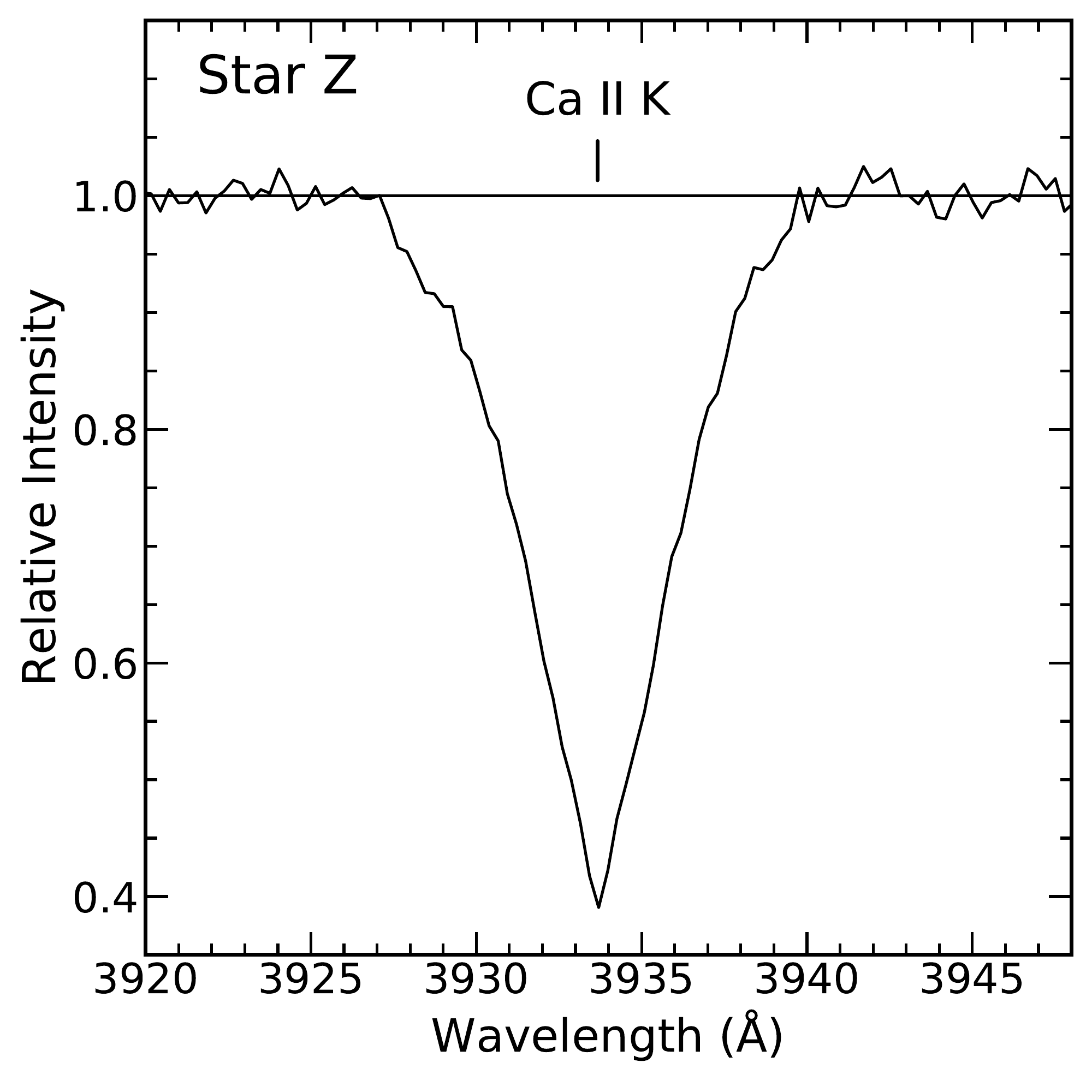}
\caption{Spectra covering the \ion{Ca}{2} line region for Stars X and Z.
   }
\label{CaII_spectra}
\end{figure}

Star X's spectrum covering the \ion{Ca}{2} K line at 3933.66 \AA \ is shown in
Figure~\ref{CaII_spectra}. As expected due to interstellar grain destruction
releasing refractory Ca atoms following SN shock passage \citep{RS1952}, the
strength of the high-velocity \ion{Ca}{2} absorption features are noticeably
stronger here relative to the low-velocity interstellar features compared to
that seen for \ion{Na}{1} (Fig.~\ref{NaI_spectra}).

The spectrum of Star Y with a projected location just west of the remnant's
bright nebula NGC~6992 also showed red and blue high-velocity interstellar
components indicating it too must lie behind the remnant
(Fig.~\ref{NaI_spectra}; middle panels). Components with approximate LSR
velocities of $-50, +50$, \& +95 km s$^{-1}$ are seen for both D1 and D2 lines.

With a Gaia DR2 reported distance nearly identical to that of Star X, namely  $735
\pm 25$ pc, Star Y's spectrum strengthens a maximum distance limit to the
backside of the Cygnus Loop of  $\simeq$760 pc.  However, since both stars are
located near the remnant's eastern limb, a distance limit to the centre of the
remnant must take into account the spherical curvature and any aspherical
geometry of the Cygnus Loop.

Lastly, the spectrum of Star Z, a 10.7 mag late A/early F star located near the
remnant's north-central limb, showed the broadest and most complicated system
of \ion{Na}{1} absorption lines (Fig.~\ref{NaI_spectra}; lower panels). At least five
high-velocity \ion{Na}{1} D1 and D2 absorption features are detected with
velocities ranging from $-158\pm 5$ to $+243\pm 5$ km s$^{-1}$. Both for
clarity and limited by our spectral resolution, we have chosen not to show
deconvolved components for the overlapping region in between the D1 and D2 lines. 
At Star Z's estimated distance of $864 \pm 30$
pc, the detection of these many blue and red  shifted \ion{Na}{1} D1 and D2
absorption features is consistent with a maximum backside remnant distance
limit of 760 pc set by Stars X and Y.

A spectrum of Star Z covering the \ion{Ca}{2} K line is shown in
Figure~\ref{CaII_spectra}. Although our spectral resolution and S/N at 3934 \AA \
is too low to resolve the expected individual high-velocity
interstellar Ca components from the stellar \ion{Ca}{2} absorption profile,
there are hints of some components along the edges of the profile's blue and
red wings.  

\bigskip

\section{Discussion}

\subsection{The Cygnus Loop's Distance}

A previous search for high-velocity components associated with the remnant
conducted by \citet{Welsh2002} likely failed due to their selection of stars
with distances less than $\sim$700 pc. However,  spectra of stars with distance
in excess of 700 pc presented in Figures~\ref{NaI_spectra} and
\ref{CaII_spectra} show clear evidence for high-velocity interstellar
\ion{Na}{1} and \ion{Ca}{2} absorptions from the Cygnus Loop's expanding shell. 

\begin{figure*}[t]
    \begin{center}
    \leavevmode
    \includegraphics[width=0.48\linewidth]{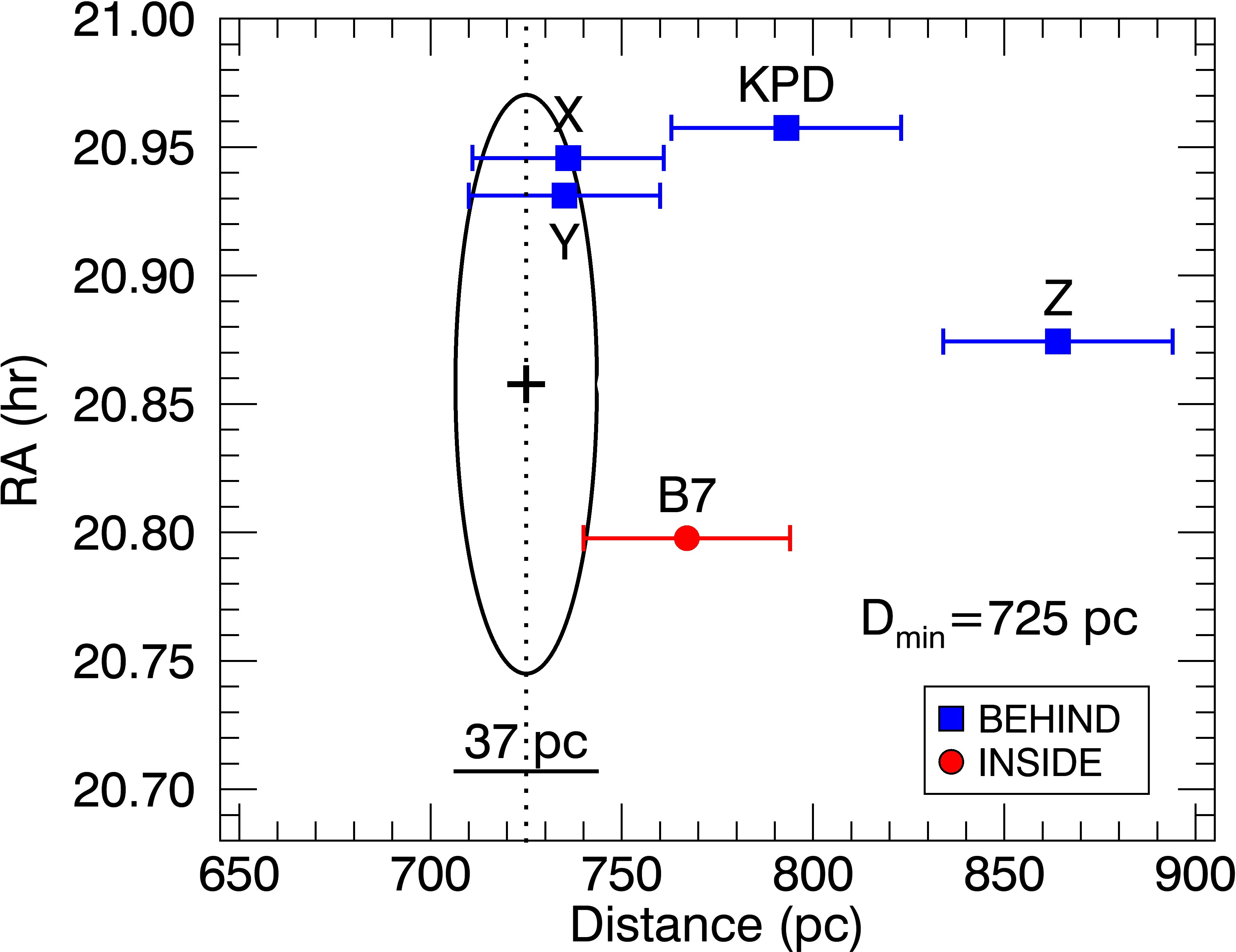}
    \includegraphics[width=0.48\linewidth]{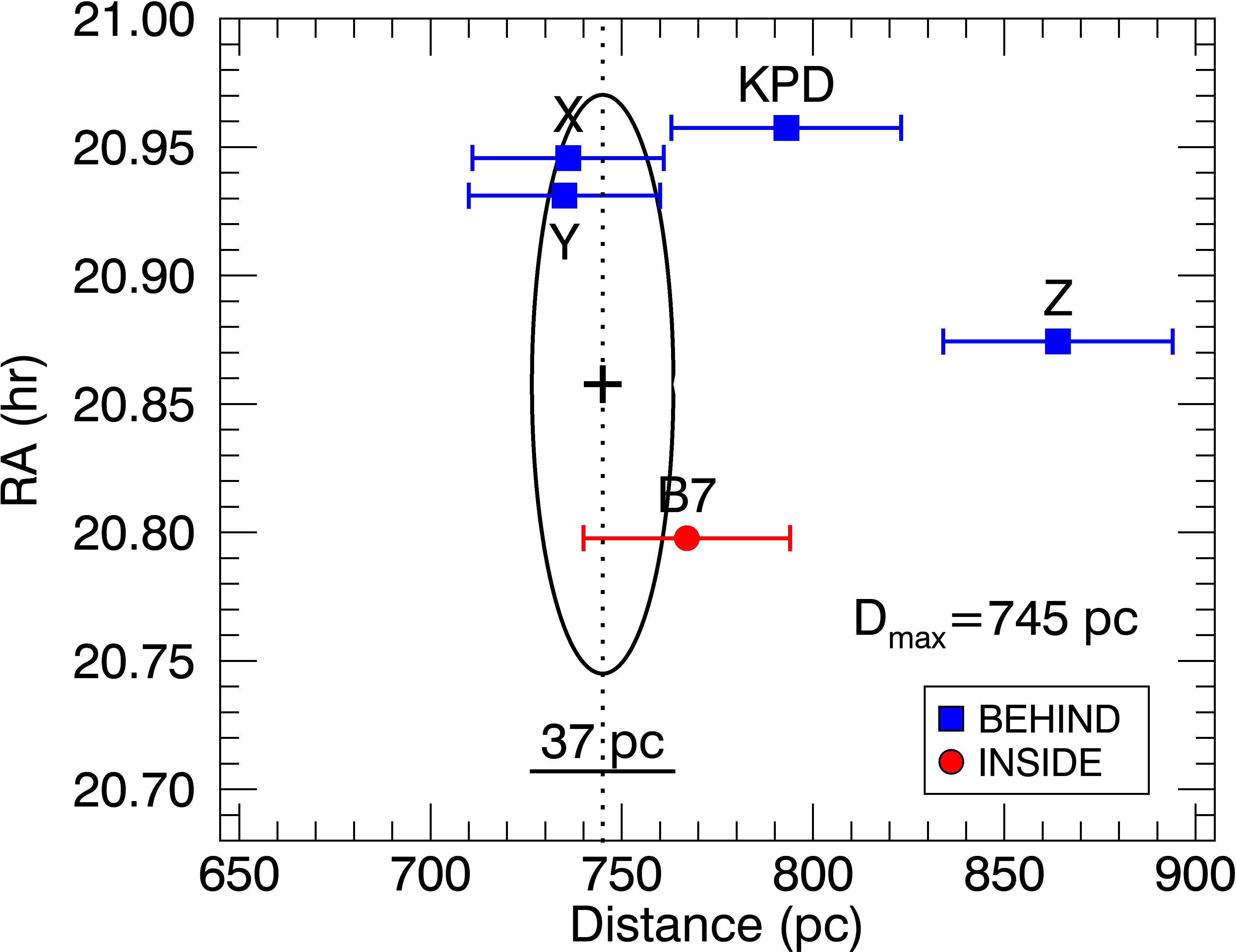} \\
    \includegraphics[width=0.48\linewidth]{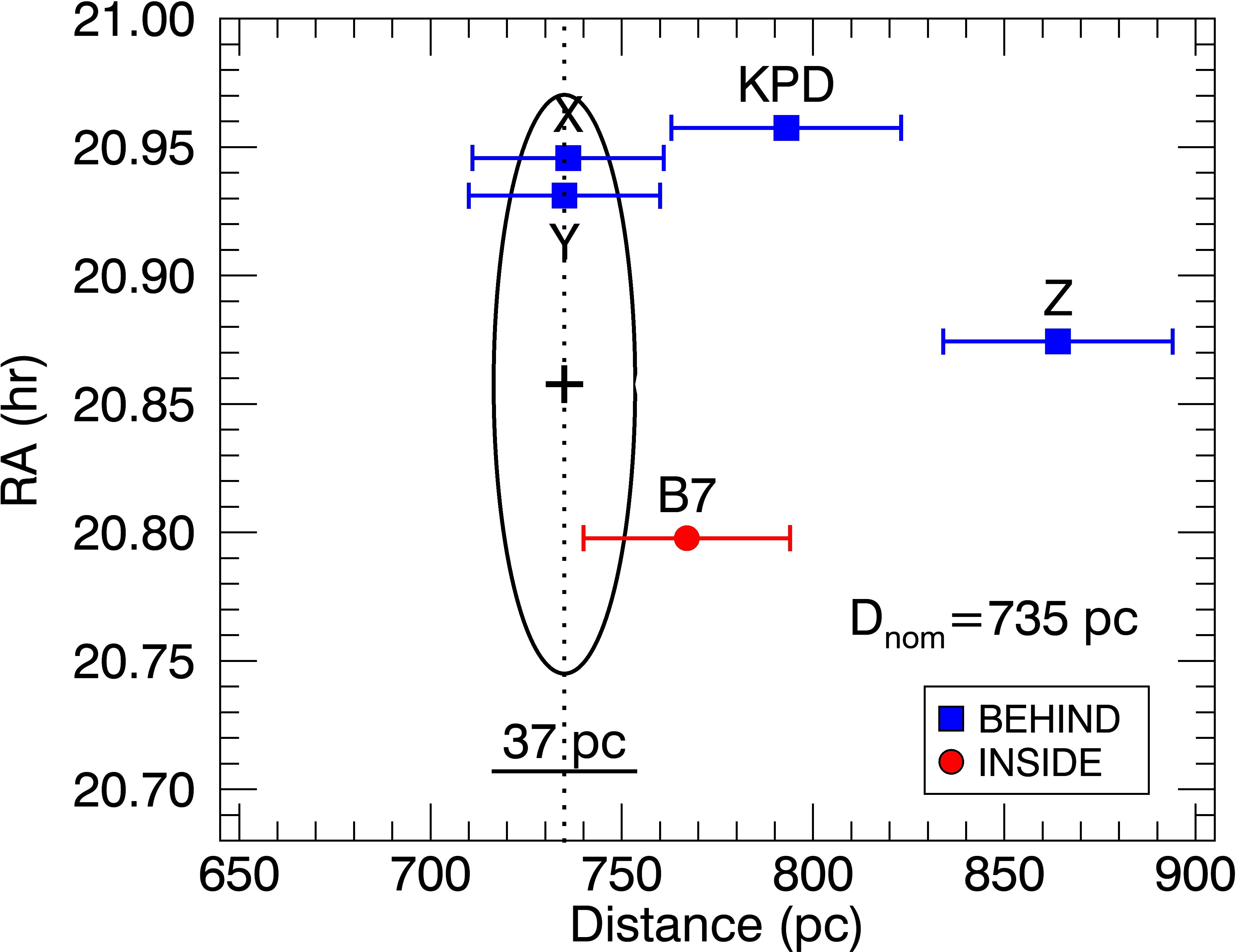}
    \includegraphics[width=0.48\linewidth]{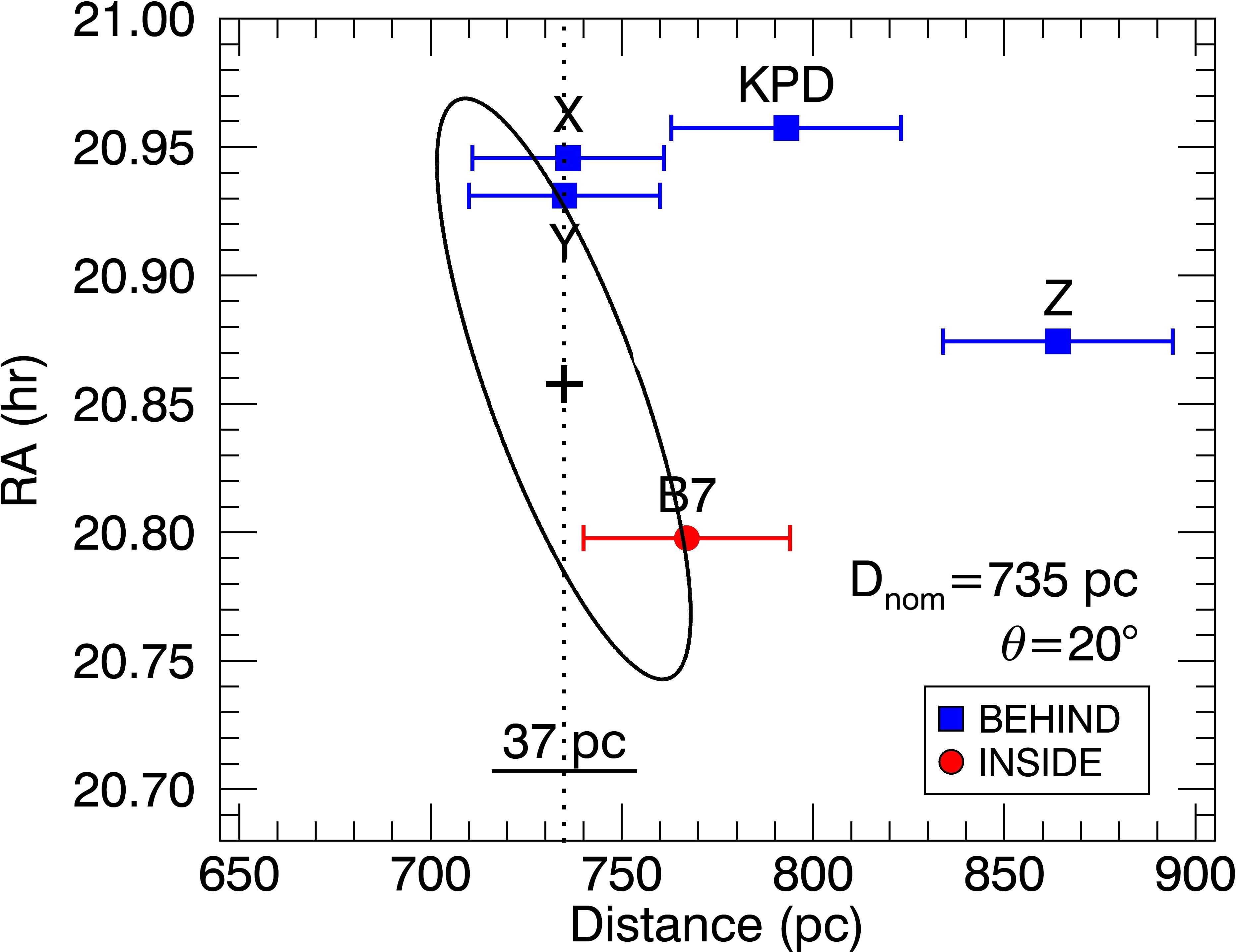} 
\caption{Plots of Gaia distances vs.\ RA  for Stars X, Y, and Z plus KPD
and the B7 star with the ellipse representing the slightly aspherical remnant.
The blue squares represent stars behind the remnant, while the red circle
represents the B7 star that resides inside the remnant. The cross represents
the location of the centre of the remnant.  {\bf{Upper Left:}} Minimum distance to the
remnant while having the B7 star located inside the remnant.  {\bf{Upper Right:}}
Maximum distance to the remnant while having Star Y located behind the remnant.
{\bf{Lower Left:}} Nominal distance to the remnant from the average of the minimum and
maximum distances.  {\bf{Lower Right:}} Nominal distance with a tilted oblate spheroid
geometry for the remnant, such that Stars X and Y are located behind the
remnant while the B7 star is located inside the remnant at their nominal
distances.  }

\label{distance_plots}
   \end{center}
\end{figure*}

In addition to the relatively high red and blue radial velocities detected for
\ion{Na}{1} lines in the spectra of Stars X, Y, and Z, the
decreased ratio of \ion{Na}{1} D1/\ion{Ca}{2} K equivalent widths in Star X is 
consistent with these components arising from shocked gas and subsequent grain
destruction in accord with the well-established Routly-Spitzer effect
\citep{RS1952,Vall1993}.  Whereas the D1/K EW ratio is 1.3 for the low velocity
component in the spectrum of Star X, the D1/K ratio is only 0.36 and 0.17 for
the $-60$ and $+95$ km s$^{-1}$ components.  Although we did not measure the \ion{Ca}{2}
equivalent widths for Stars Y and Z, their high-velocity \ion{Na}{1} components are as
large or larger than that seen in Star X and hence would be expected to
exhibit strong high-velocity \ion{Ca}{2} absorptions with corresponding low
D1/K equivalent width ratios.

\begin{figure*}[t]
    \begin{center}
    \leavevmode
    \includegraphics[width=0.48\linewidth]{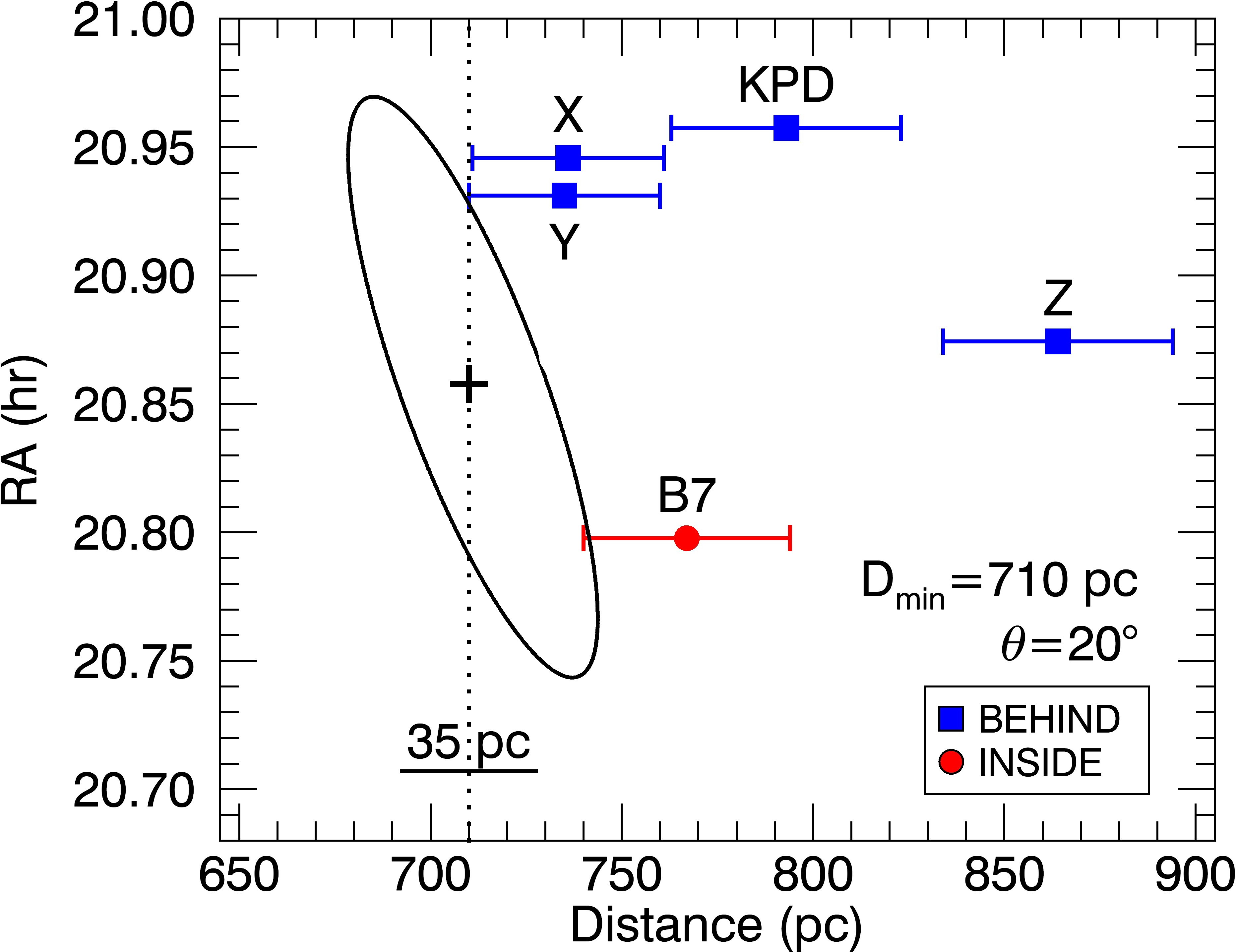}
    \includegraphics[width=0.48\linewidth]{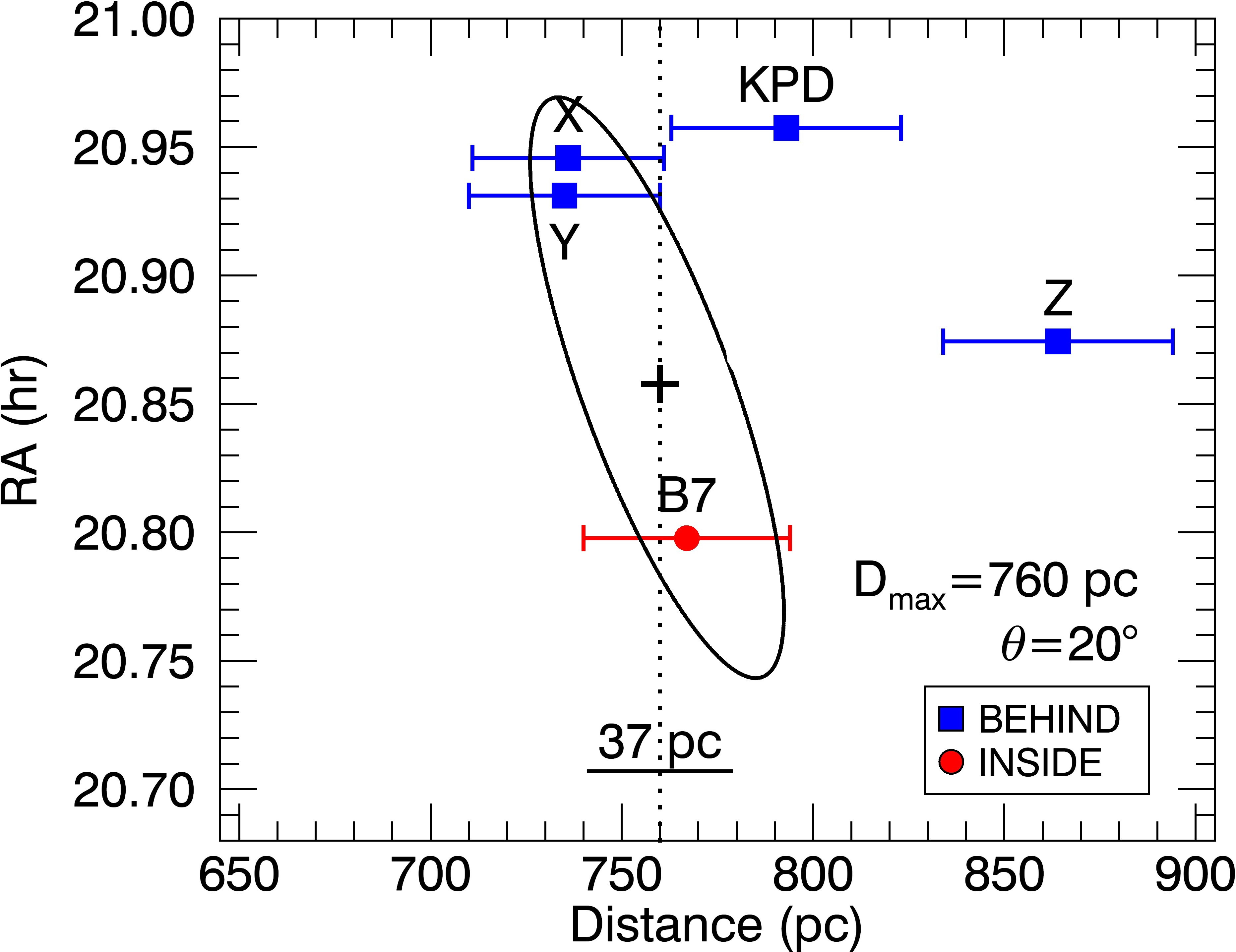} 
\caption{Similar plots to those shown in Figure~\ref{distance_plots}. {\bf{Left
Panel:}} Minimum distance of 710 pc to the remnant for a 20 degree tilt such that
the B7 remains inside the remnant at its extrema distance while Stars X and Y
are behind the remnant.  {\bf{Right Panel:}} Maximum distance of 760 pc to the remnant
for 20 degree tilt such that Stars X and Y lie behind the remnant at their
extrema distances while the B7 star lies within the remnant.  }

\label{more_distance_plots}
   \end{center}
\end{figure*}

The detection of both blue and red high-velocity interstellar absorption
features in the spectra of Stars X and Y with Gaia estimated distances of $735
\pm25$ pc, together with several blue and red shifted interstellar absorptions
in Star Z located at 864 pc, place a firm maximum distance limit to the
backside of the Cygnus Loop of 760 pc.  This is consistent with the KPD star's
$793 \pm 30$ pc distance which showed remnant related line absorptions in its
UV spectrum implying a location behind the remnant \citep{Blair2009}.

Formally, these detections allow for all distances less than 760 pc for the
backside of the remnant including the $\simeq 550$ pc estimate proposed by
\cite{Blair2009}. However, if the stellar wind from the B7 star, BD+31 4224,
located along the remnant's northwestern limb is truly interacting with the
remnant's expanding shock wave as argued by \citet{Fesen2018}, then its Gaia
estimated distance of $767 \pm 27$ pc places severe constrains on the remnant's
distance. 

Adopting the premise that the B7 star is interacting with the remnant, this
effectively anchors the Cygnus Loop to distances within the B7 star's parallax
measurement errors; namely, the remnant's rear hemisphere can be no closer than
740 pc, and its frontside can be no further than 794 pc.  At such distances, the
Cygnus Loop's main shell diameter of 2.85 degrees implies a linear diameter of
$\simeq$ 37 pc. 

The combination of the B7 star interacting with the remnant's northwestern
hemisphere and the requirement that both Stars X and Y at 735 pc must lie
behind the remnant places additional constrains on the maximum and minimum
distances to the Cygnus Loop's centre of expansion.  Measured from its optical
emission structure, the remnant is slightly asymmetric corresponding to an
ellipse in the distance verses RA plane.  Figure~\ref{distance_plots} shows
plots of Gaia distances verses RA for Stars X, Y, Z, plus KPD, and the B7 star.
Stars that must lie behind the remnant are shown as blue squares, while the B7 star
which must be inside or interacting with the backside of the remnant is shown
as a red circle.  The projection of the remnant, centred at 20.854 hours RA
(J2000: 20:51:14), in distance vs.  RA is illustrated in these four plots.

The top left panel of Figure~\ref{distance_plots} shows that an aspherical
remnant at a distance of 725 pc is just barely able to intersect the B7 star at it's
Gaia shortest estimated distance, thereby setting a minimum
distance to the Cygnus Loop's centre at 725 pc. In similar fashion, the top
right panel shows that this same 37 pc diameter remnant cannot be centred any
farther away than $\simeq$745 pc so as to have Stars X and Y behind the
remnant's rear shell at their maximum Gaia estimated distances.  

The fortuitous situation of having two stars, namely Stars X and Y, located
behind the remnant's eastern limb at virtually the same distance (735 pc)
suggests the remnant's rear eastern hemisphere is likely to be close to this
735 pc value. However, no Cygnus Loop distance between the 725 and 745 pc
limits discussed above can simultaneously fulfill the requirements that Stars X
and Y lie behind a 37 pc diameter remnant at their nominal Gaia 735 pc value
while still having the B7 star inside or just touching the remnant's rear NW
shell at its nominal 767 pc distance.  This is illustrated in the lower left
panel.

A solution is that the Cygnus Loop's main northern shell containing its
brightest nebulae is aspherical and tilted.  As shown in the lower right panel
of Figure~\ref{distance_plots}, if the remnant's eastern half is tilted toward
us then the various stellar distances would be in better agreement; that is,
the B7 star could lie inside the remnant's NW limb while having Stars X and Y
behind the remnant's eastern limb.

If true, this might imply the remnant's eastern limb expanded more rapidly
towards us. Consequently the remnant's western limb, particularly its
northwestern limb, is located farther away.  Specifically, an asymmetric
remnant with a diameter $\simeq$ 37 pc tilted some 20 degrees allows Stars X
and Y to lie behind the remnant's eastern limb while still having the B7 star
within the remnant.  A mild asymmetry along our line of sight would be
consistent with {\it{Suzaku}} and {\it{XMM-Newton}} X-ray studies of the
remnant \citep{Uchida2009}.

\begin{deluxetable*}{lcl}
\tablecolumns{3}
\tablecaption{Fundamental Properties of the Cygnus Loop}
\tablewidth{0pt}
\tablehead{ \colhead{Property}   &  \colhead{Value} & \colhead{Details \&  References} }
\startdata
{\bf Distance and Size}             &                      &                                                                        \\
Distance                     & 735$\pm$25 pc        & based on Gaia DR2 parallax measurements of several stars: this paper \\
Diameter                     & E-W 37$\pm$2 pc      & adopting a 735 pc distance and an E-W diameter of 2.88 degrees: this paper  \\
                             & N-S 47$\pm$3 pc      & adopting a 735 pc distance and a N-S diameter of 3.65 degrees: this paper  \\
                             &                      &                                               \\
{\bf Blast Wave}             &                      &                                                                        \\
n$_0$                        & 0.4$\pm$0.1 $\rm cm^{-3}$  & from Balmer filaments and optical depth of \ion{O}{6}: \citet{Raymond2003}    \\
Proper Motion                & $0.100\pm$0.025$^{\prime\prime}$ yr$^{-1}$ & mean value for 18 positions: \citet{Salvesen2009}  \\
V$_{\rm shock}$              & 348$\pm$87 $\rm km~s^{-1}$ & combining  $735$ pc distance and Balmer filament proper motions: this paper \\
                             &                      & $343 \pm26$ km s$^{-1}$: \citet{Medina2014}; $360$ km s$^{-1}$: \citet{Raymond2015} \\
                             &                     &                                               \\
{\bf Sedov}                  &                      &                                           \\
Age                          & 21,000$\pm$4,000 yr  & t = 0.4$\times$radius/$348\pm 87$ km s$^{-1}$: this paper \\
E                            & $0.7\pm0.2 \times 10^{51}$ erg      &   n$_0$ = $0.4 \pm 0.1$ cm$^{-3}$; 
                              radius  = $1.54\times10^{19}$ cm E$_{51}^{1/5}$ n$_{0}^{-1/5}$ t$_{\rm 1000~yr}^{2/5}$ \\
V$_{\rm shock}$              & 367$\pm$80 km s$^{-1}$  & V = 1950 km s$^{-1}$ E$_{51}^{1/5}$ n$_{0}^{-1/5}$ t$_{\rm 1000~yr}^{-3/5}$  \\
                             &                      &                                               \\
{\bf Transitioning Shock}       &                      &                                               \\
n$_0$                        & 1.5$\pm$0.5 $\rm cm^{-3}$  & \citet{Long1992}                        \\
Proper Motion                & 0.070$\pm$0.008$^{\prime\prime}$ yr$^{-1}$ & northeastern Balmer filament transitioning to radiative: \citet{Blair2005}  \\
V$_{\rm shock}$              & 244$\pm$28 $\rm km~s^{-1}$ & combining 735 pc distance \& filament proper motion: this paper \\
                             &                     &                                               \\
{\bf Cloud Shock}            &                      &                                               \\
n$_0$                        & $\simeq$ 3 - 10 cm$^{-3}$  & \citet{Raymond1988}; \citet{Levenson1998}; \citet{Leahy2003}; \\
                             &                     &        \citet{McE2011} \\

Proper Motion                & 0.030 - 0.040$^{\prime\prime}$ yr$^{-1}$      & \citet{Hubble1937,Shull1991}           \\
V$_{\rm shock}$              & 105 - 140  $\rm km~s^{-1}$ & combining 735 pc distance and filament proper motion: this paper  \\
                             &                              & \citet{Minkowski1958} \&  \citet{Doro1970} find 110 to 115 $\rm km~s^{-1}$ for bright filaments \\
\enddata
\end{deluxetable*}

Although larger asymmetries or tilts are possible, the remnant's fairly
circular projected morphology in its main shell of emission suggests large
departures from symmetry are unlikely. Thus, adopting an effective
maximum tilt of 20 degrees, the plots shown in Figure~\ref{more_distance_plots}
suggest more realistic distance limits to the remnant. In the left panel, a 35 pc
diameter remnant tilted 20 degrees whose east and west limbs match the observed
Cygnus Loop's dimensions cannot be closer than 710 pc and still have the B7
star in physical contact with the remnant.  A maximum distance for a 20 degree
tilt of 760 pc is shown in the right panel of this figure, where Star Y is at
its maximum distance while still lying behind the remnant.

Consequently, from the analyses discussed above, we conclude the centre of the
main, northern portion of the Cygnus Loop containing the majority of the
remnant's bright optical nebulae lies at a distance of $735 \pm 25$ pc with
distances between 735 and 745 pc slightly favored based on the B7 star's
nominal 767 pc distance.  Interestingly, this distance estimate is only a bit
less than the 770 pc value proposed by Minkowski some 60 years ago.  At 735 pc,
the remnant's angular diameter of 2.88 degrees corresponds to $\simeq$37 pc
with an uncertainty of just a few parsecs due to the remnant's degree of
asphericity.

\begin{figure*}[t]
    \begin{center}
    \leavevmode
     \includegraphics[width=0.95\linewidth]{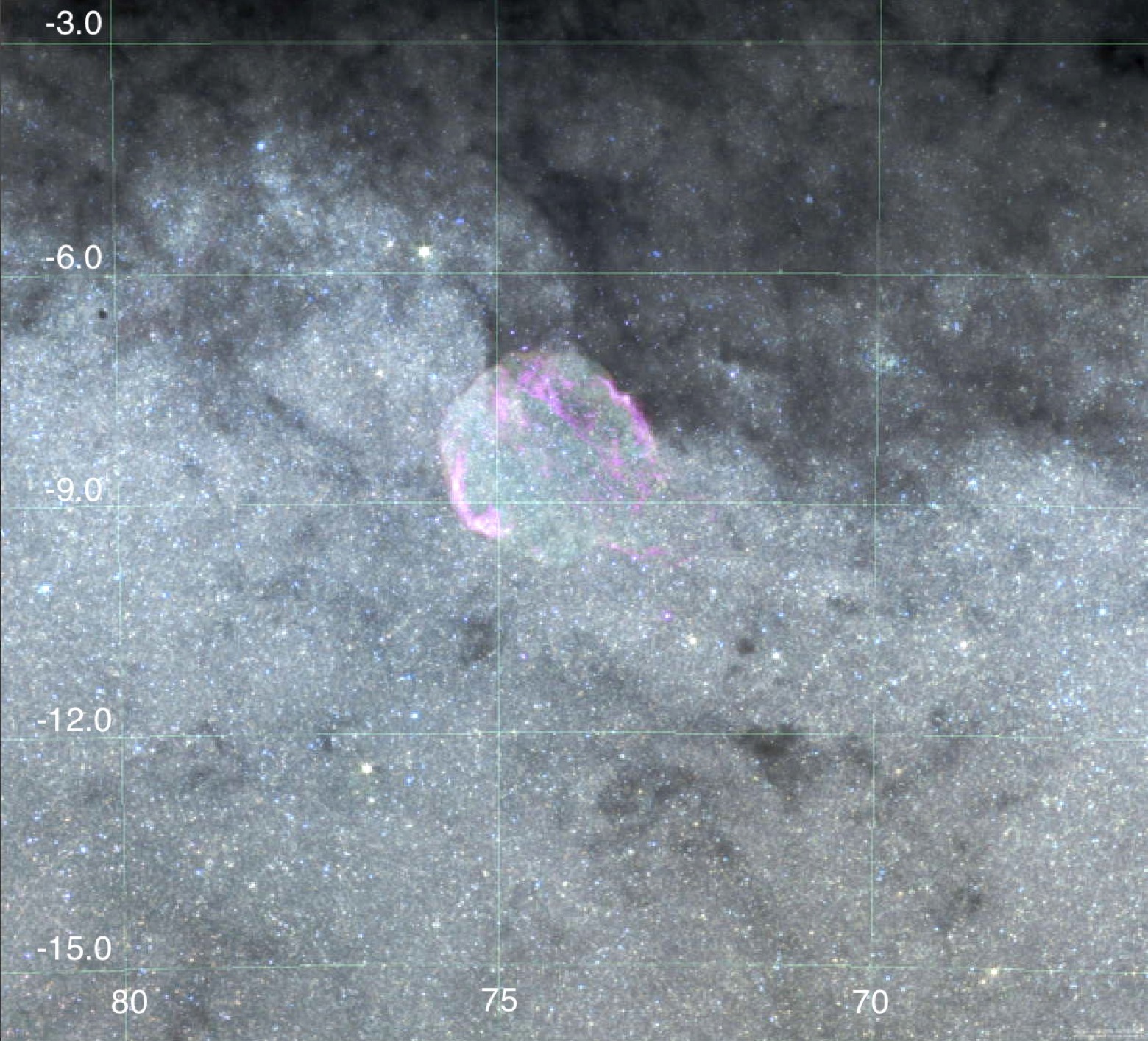} \\
\caption{A wide-angle composite image made from optical Mellinger RGB images, Planck
857 GHz infrared data (gray), {\sl GALEX} UV images (pink) and {\sl ROSAT} X-ray images (white)
of the Milky Way region around the Cygnus Loop. Galactic coordinates are
shown. The remnant's brightest optical and X-ray emission features can be seen
immediately adjacent to local dust clouds.
}
\label{dust_clouds}
   \end{center}
\end{figure*}

\subsection{The Cygnus Loop's Fundamental Properties}

Assuming our distance estimate of 735 pc to the remnant's centre and its
east-west diameter of 37 pc, we list in Table 3 several basic  properties of
the Cygnus Loop including shock velocities in low density regions, in slightly
denser regions where the shock is transitioning from non-radiative Balmer
dominated filament emission to radiative, and finally in denser cloud regions.
As seen in this table, we find good agreement with spectroscopically inferred
shock velocities for the three different density regions using the new distance
estimate. 

A distance of 735 pc implies a lower SN energy of $0.7 \pm 0.2 \times 10^{51}$
erg than the canonical SN explosion energy of $1 {-} 2 \times 10^{51}$ erg.
This value is similar to earlier estimates of E$_0$ = $3 {-} 7 \times 10^{50}$
(d$_{\rm pc}$/770)$^{5/2}$ erg using Minkowski's distance of 770 pc
\citep{Rapp1974,Falle1982,Ballet1984,MT1999}.  However, these previous
estimates, as well as ours, do not include the remnant's southern breakout
region which would raise the $E_{0}$ value a bit, possibly close to $1 \times
10^{51}$ erg (see discussion in $\S$4.4).

\subsection{The Remnant's Environment Driven Morphology}

An improved knowledge about the Cygnus Loop's distance can help our
understanding about its local environment and how this has impacted the
remnant's gross morphology and emission structure.  Below, we briefly review
past conclusions regarding the remnant's interstellar environment and
evolution, specifically its proposed origin as a progenitor wind-blow cavity
explosion. We then present multi-wavelength emission maps that offer a new look
into the remnant's interstellar environment as the origin of the remnant's
large-scale structure.

\subsubsection{A Historical Prospective}

As seen across many wavelengths, the Cygnus Loop's overall morphology is far
from that expected for an expanding sphere in a uniform medium.  Its northern
portion is dominated by bright and nearly opposing nebulae, NGC~6960 in the
west and NGC~6992 \& NGC~6995 in the east (see Fig.\ 1). In addition, there is
an extensive complex of emission filaments along it north-central region
(Pickering's Triangle) with numerous other smaller nebulae and filaments, all
surrounded by a nearly complete X-ray emitting and Balmer-dominated shock front
shell. 

Fainter and much sparser optical, X-ray, and radio emissions extending some 90
arcminutes to the south from the remnant's nominal centre mark a prominent
rupture-like structure best seen in X-rays \citep{Levenson1997,Aschen1999}.
This `blowout region' exhibits such different radio polarization properties
that it has led some to propose it to be a separate SNR
\citep{Uyan2002,Sun2006,West2016}.

The idea that the Cygnus Loop is a progenitor cavity explosion goes back to the
earliest X-ray studies of the remnant.  The discovery by \citet{Tucker1971}
that the remnant's X-ray inferred blast wave velocity was almost three times
the velocity of the optical filaments, led \citet{McKee1975} to suggest the
remnant's bright optical nebulae are due to dense interstellar clouds recently
hit by the blast wave. 

This notion was subsequently questioned by \citet{McCray1979},
\citet{Charles1985}, and \citet{Braun1986} who viewed such a scenario as
unlikely to have created `such a nicely spherical shell'. They instead
suggested that the Cygnus Loop's shape was the result of a SN explosion taking
place in a largely spherical, wind-driven cavity pre-dating the SNR, produced
by the SN progenitor itself.    

\begin{figure*}[t]
    \begin{center}
    \leavevmode
    \includegraphics[width=0.95\linewidth]{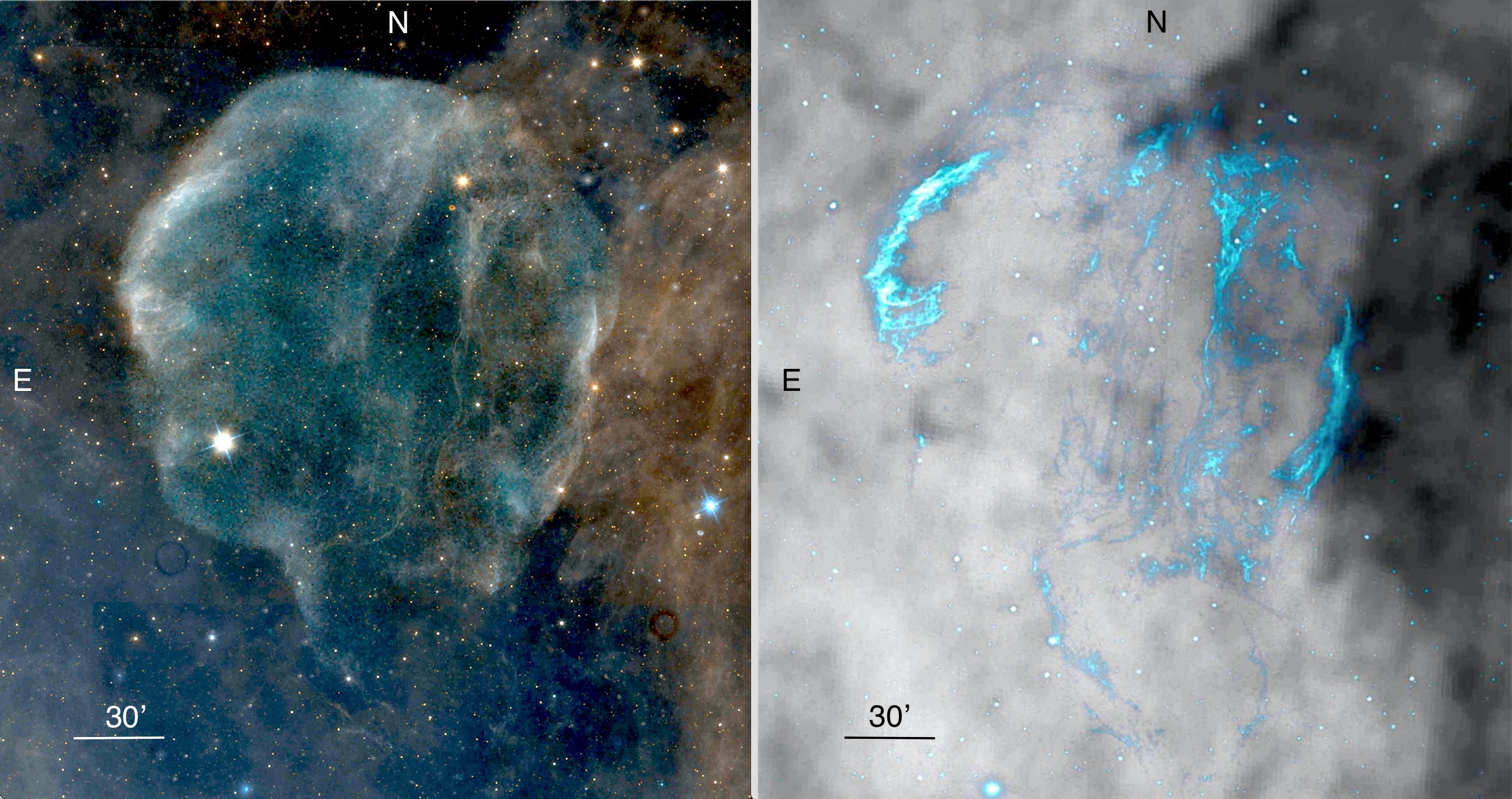}
\caption{
{\bf{Left Panel:}} Composite positive image made from {\sl ROSAT} X-ray (blue), {\sl GALEX} UV (white), and {\sl WISE}  12 \& 22 $\mu$m infrared data
(blue and red) highlighting the positional coincidence of the Cygnus Loop's X-ray and UV emissions with local dust clouds, and the
remnant's expansion northward toward a region relatively clear of interstellar clouds. 
{\bf{Right Panel:}} Composite image
made from Planck 857 GHz (gray) data and {\sl GALEX} UV data (positive blue) showing clouds all along 
the Cygnus Loop's eastern nebulae, plus a small cloud one degree south of NGC~6995 coincident 
with the remnant's small isolated southeastern emission knot.
 }
\label{WISE_n_UV}
   \end{center}
\end{figure*}

There has been considerable support for the Cygnus Loop being a cavity
explosion remnant.  Spectral X-ray analyses revealed evidence for a high-mass
progenitor through observed enrichments of O, Ne, and Mg suggesting ejecta from
a $\simeq$15 solar mass core-collapse SN.  In addition, numerous model
investigations have concluded that the remnant's properties are consistent with
a SN explosion inside an interstellar cavity created by the progenitor star
\citep{Hester1994,Graham1995,Levenson1998,Uchida2009}.  Recently,
\citet{Fang2017} conducted 3D hydrodynamic simulations assuming an anisotropic
and latitude-dependent progenitor wind and achieved good agreement with the
remnant's observed morphology.   

However, its been known since the early 1920's, even before it was realized to
be a SN remnant, that the Cygnus Loop's bright western nebula, NGC~6960,
marks the boundary of a significant drop off in star density farther
to the west due to an adjacent interstellar cloud
\citep{Wolf1923,Duncan1923,Oort1946}. This difference is now understood as dust
destruction by the remnant's blast wave.  Obscuration
caused by this cloud is readily apparent on photographic images
of NGC~6960 such as those published by \citet{Ross1931}.  

Extinction differences around NGC~6960 have been studied by \citet{Chamber1953}
and \citet{Bok1957}, with H~I and CO emission maps of the region showing
a good correlation with the observed optical obscuration
\citep{deNoyer1975,Scoville1977}.  While an analysis of mid-infrared images of
the remnant and its local interstellar environment led \citet{Arendt1992} to
suggest no direct link between this western cloud and NGC~6960, a study
of X-ray and optical emissions along the Cygnus Loop's western limb led
\citet{Levenson1996,Levenson1998} to conclude the remnant was, in fact,
directly interacting with this cloud.

Observations revealing an expanding shell of neutral hydrogen along the Cygnus
Loop's northeastern limb led \citet{Leahy2003} to suggest the presence of an
interstellar cloud in this region which was given a small outward velocity by
the Cygnus Loop's progenitor star's ionizing radiation. A similar analysis was
subsequently proposed for the remnant's southeastern limb  \citep{Leahy2005}.
In both regions, interstellar clouds were proposed to be immediately adjacent
to the remnant.  

From a deep optical survey of the remnant's emission structure,
\citet{Levenson1998} concluded the remnant's brightest optical emissions trace
interactions of its blast wave with dense interstellar clumps. They found faint
nonradiative, Balmer-dominated shock emission filaments forming a nearly
circular and complete shell around the remnant's northern region with a radius
of 1.9 degrees.  They argued that the remnant's bright optical emission arising
from radiative shocks are only seen where the blast wave encounters denser
inhomogeneities in the local interstellar medium, and it is these interactions
which dominate the remnant's global appearance and evolution.   

\subsubsection{A Multi-wavelength Analysis}

In order to investigate the interstellar environment around the Cygnus Loop, we
constructed a series of multi-wavelength composite images of its neighboring
regions using on-line data archives. These included Planck infrared
images \citep{Planck2014}, Wide-field Infrared Survey Explorer ({\sl WISE}) data \citep{Wright2010},
{\sl ROSAT} HRI and PSPC X-ray images, optical survey images including the Digital
Sky Survey (DSS2) and the Mellinger survey \citep{Mell2009}, and near UV image
data from the Galaxy Evolution Explorer ({\sl GALEX}) \citep{Morr2007}.

Figure~\ref{dust_clouds} shows a roughly $12 \times 15$ degree wide composite
image of the Milky Way around the Cygnus Loop covering Galactic coordinates $l
= 66\degr$ to $82\degr$, $b = -2.8\degr$ to $-15.5\degr$, constructed from
Planck 857 GHz, {\sl ROSAT} PSPC, {\sl GALEX} UV, and Mellinger images.  This composite
relates the projected locations of the remnant's X-ray and optical emissions
with interstellar dust clouds.  

The image shows that the remnant, located well off the Galactic plane between
Galactic latitudes $-6.9\degr$ to $-9.8\degr$, is situated in a broad channel
largely empty of dust clouds, bounded on its western limb by a large, extended
dust cloud stretching roughly from $l = 69\degr$ to $75\degr$.  Eastern edges
of this cloud coincide with the remnant's bright western and northwestern
nebulosities, namely NGC~6960 and Pickering's Triangle.   Smaller, less dense
dust clouds near $l = 76\degr, b = -9\degr$ appear to coincide with the
remnant's eastern X-ray and optical emissions, namely NGC~6992, NGC~6995, and
IC~1340.

\begin{figure*}[t]
    \begin{center}
    \leavevmode
    \includegraphics[width=0.95\textwidth]{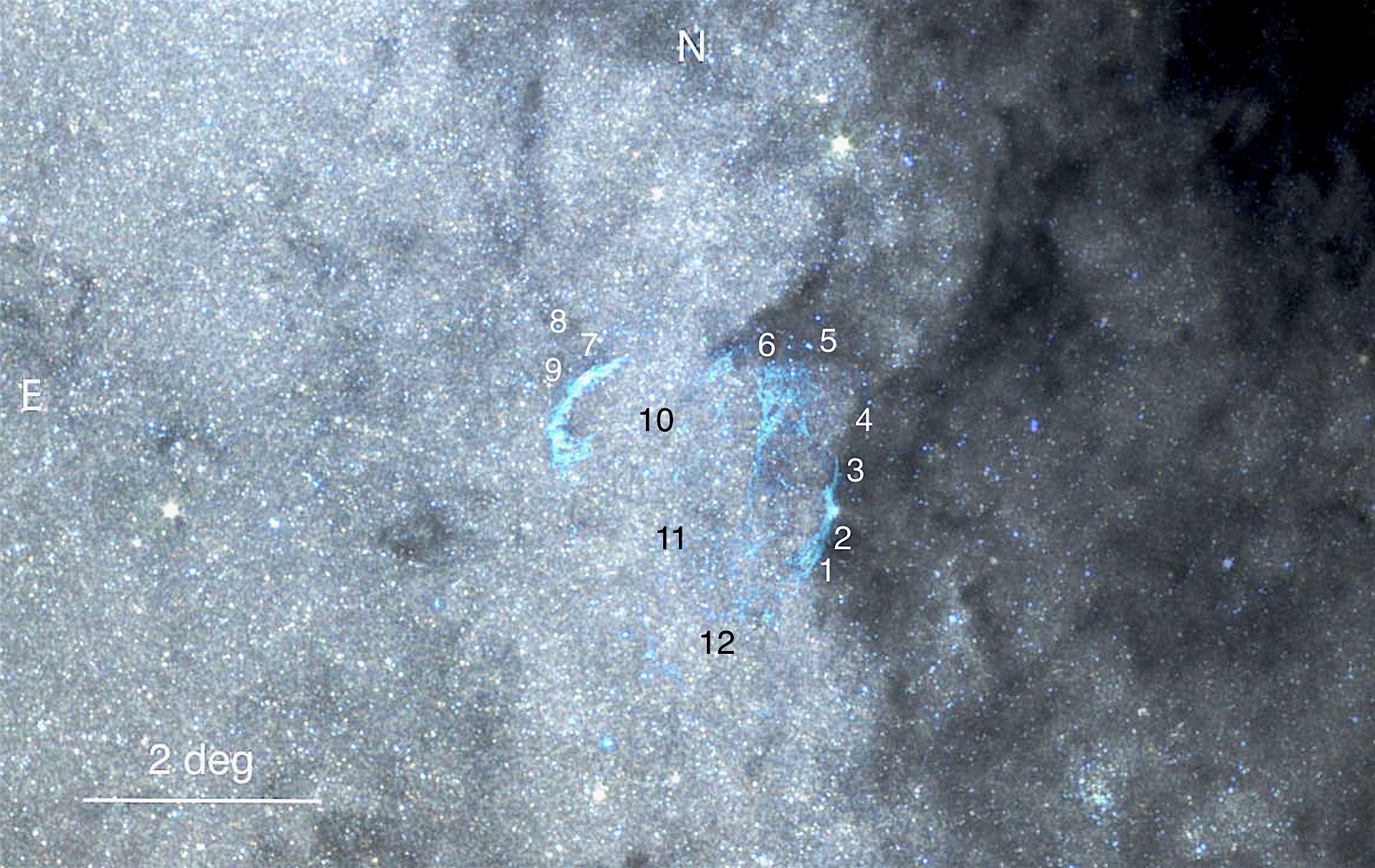}
\caption{Composite image made from optical Mellinger RGB images (white), Planck infrared (gray) and {\sl GALEX} UV (blue) data showing 
the presence of interstellar dust clouds projected along all of the remnant's bright optical and UV nebulosities. Marked are the
12 regions where reddening measurements were made using Pan-STARRS and 2MASS data on cumulative dust reddening with distance.
 }
\label{optical_n_dust}
   \end{center}
\end{figure*}

Additional composite images showing  the interstellar medium around the Cygnus Loop
using {\sl WISE}, {\sl GALEX}, {\sl ROSAT}, and Planck data are shown in Figure~\ref{WISE_n_UV},
but now in an equatorial coordinate format.  The left panel shows the remnant's {\sl ROSAT} X-ray, {\sl GALEX} UV 
and {\sl WISE} infrared emissions
in relation to projected locations of infrared emission from dust clouds. Especially notable is
the presence of dust clouds along most of the remnant's western limb. The appearance of the remnant's emission
relative to these clouds gives the impression that the remnant lies on the near side of these clouds.
This is supported by the small extension or bulge of remnant's X-ray emitting shock front to the west of the northern tip
of NGC~6960. We also note that the remnant's seemingly greater expansion to the north coincides with a 
region relatively free of significant interstellar clouds. 

The other multi-wavelength composite shown in the right panel of
Figure~\ref{WISE_n_UV} reveals excellent correlation between the remnant's
bright radiative UV and optical emission nebulae and projected locations of
dust clouds. This includes dust clouds all along the remnant's bright eastern
nebulosities, plus a small dust cloud coincident with the remnant's
southeastern emission knot roughly one degree south of NGC~6995 that has been
studied by \citet{Fesen1992}, \citet{Graham1995}, and \citet{LG2001}.

Although there is considerable evidence for direct cloud interaction along
the Cygnus Loop's western limb \citep{Levenson1996}, there has not been
previously published data pointing to specific dense clouds correlated with the remnant's
other strong X-ray/UV/optical emission features such as Pickering's Triangle,
NGC~6992 or NGC~6995.  The locations of the dust clouds seen in
Figures~\ref{dust_clouds} and \ref{WISE_n_UV} greatly clarifies where and
why the remnant looks the way it does across many wavelengths.

Using X-ray data and deep optical images covering the entire remnant,
\citet{Levenson1998} attempted to construct a three-dimensional picture of the
Cygnus Loop.  They concluded  that dense clouds and atomic gas together form
the walls of a cavity in which the SN occurred. However, the composite images
shown in Figures 6 and 7 make it clear that
the ``cavity'' is simply a region relatively free of interstellar clouds.

We agree with Levenson et al.'s suggestion that a large cloud located along the
rear of the remnant is responsible for its western limb emission since it is
consistent with our distance estimate for the remnant centre vs. that of the B7
star. However, they claim that the cloud that is the cause for the bright
northeastern emission is also on the far side, whereas we find evidence for the
remnant's eastern emissions to be closer than emission along the west and
northwestern limbs.     

One can test the idea that projected coincidences of these dust clouds do, in
fact, signal interactions of the remnant's shock by using the new distance
described above along with reddening versus distance estimates toward these
clouds. Figure~\ref{optical_n_dust} shows 12 regions for which we obtained
reddening values as a function of distance from the three-dimensional maps
presented in \citet{Green2015} which utilized Pan-STARRS \citep{Schlafly2014}
and 2MASS \citep{Skrutskie2006}
photometry\footnote{http://argonaut.skymaps.info}.  Figure~\ref{optical_n_dust}
also highlights again the projected locations of the Cygnus Loop's optical and
UV bright nebulae relative to interstellar dust clouds and its position inside
a broad N-S channel largely empty of interstellar clouds.  

\begin{deluxetable*}{lccccccccc} \tablecolumns{10} \tablecaption{Reddening
E(B-V) Values vs. Distance for Dust Clouds Around the Cygnus Loop}
\tablewidth{0pt} \tablehead{\colhead{Region} & \colhead{Number} & \colhead{RA}
& \colhead{Dec} & \colhead{450 pc} & \colhead{550 pc} & \colhead{650 pc} &
\colhead{750 pc} & \colhead{850 pc} & \colhead{950 pc} } \startdata 
Western Molecular Cloud & 1 & 20:45:39.2 & +30:11:50 & 0.06 & 0.06 & 0.07 & 0.08 & 0.21 & 0.41 \\
 & 2 & 20:45:22.5 & +30:31:45 & 0.03 & 0.04 & 0.08 & 0.32 & 0.42 & 0.42 \\
 & 3 & 20:45:02.0 & +31:02:30 & 0.07 & 0.10 & 0.11 & 0.13 & 0.25 & 0.46 \\
 & 4 & 20:44:39.3 & +31:28:15 & 0.05 & 0.05 & 0.10 & 0.34 & 0.43 & 0.45 \\
North West Clouds & 5 & 20:46:46.5 & +32:11:45 & 0.08 & 0.13 & 0.21 & 0.21 & 0.25 & 0.26 \\
 & 6 & 20:48:24.1 & +32:09:15 & 0.04 & 0.11 & 0.19 & 0.25
& 0.29 & 0.30 \\ North East Clouds       & 7 & 20:55:51.0 & +32:00:00 & 0.06 &
0.08 & 0.09 & 0.11 & 0.13 & 0.14 \\ & 8 & 20:57:18.0 & +32:13:30 & 0.09 & 0.10
& 0.12 & 0.14 & 0.16 & 0.17 \\ & 9 & 20:57:24.0 & +31:56:30 & 0.05 & 0.07 &
0.08 & 0.10 & 0.12 & 0.15 \\  Central Regions     &10 & 20:53:10.0 &
+31:25:00 & 0.05 & 0.06 & 0.07 & 0.08 & 0.08 & 0.09 \\ &11 & 20:53:44.0 &
+30:22:00 & 0.04 & 0.05 & 0.05 & 0.06 & 0.06 & 0.07 \\ &12 & 20:50:59.0 &
+29:25:00 & 0.06 & 0.06 & 0.06 & 0.07 & 0.07 & 0.07 \\ 
\enddata
\end{deluxetable*}

Measured $E(B-V)$ values for 450, 550, 650, 750, 850, and 950 pc for the 12
regions marked in  Figure~\ref{optical_n_dust} are listed in Table 4.  The
sharp rise of extinction around 650 to 750 pc along the remnant's western limb
(Regions 1-4) is presumably due to the molecular dust cloud the remnant is
physically encountering. This rise is consistent with our derived $735 \pm 25$
pc distance.  In like fashion, the reddening measurements to the northwestern
clouds (Regions 5 \& 6) and the fainter northeastern clouds (Regions 7-9) are
also consistent with distances in excess of $\simeq$ 550 pc. In contrast, the
reddening measurements taken for projected regions toward the centre of remnant
(Regions 10-12) show reddening values that only slowly increase with distance
out to 950 pc. 

It should be noted that these reddening measurements support the view that the
B7 star lies inside the remnant as proposed by \citet{Fesen2018}.  Moreover,
since reddening measurements for all 12 locations are found to be relatively
low ($<$ 0.13) out to 550 pc,  distances to the Cygnus Loop of less than
$\simeq$ 600 pc are now firmly excluded.  Finally, as mentioned above and
discussed by \citet{Fesen2018}, the presence of some of the remnant's shock
filaments up to 30$'$ farther to the west of NGC~6960 (see Fig.\ 4 in
\citealt{Fesen1992}) suggests the remnant lies close to the near side of the
western molecular cloud.

Interaction of the remnant's blast wave with local clouds along its eastern
limb appears likely to have occurred relatively recently. \citet{Hester1986}
used the presence of optical [\ion{O}{3}] line emission behind a Balmer
filament along the remnant's eastern limb to conclude that the remnant's shock
hit a wall of denser gas a few hundred years ago. \citet{Shull1991}
cited X-ray emission just outside the remnant's eastern optical emission
filaments to suggest that the cloud was encountered less than 1000 yr ago.

In summary, we find that the long suspected SN progenitor generated cavity
walls appear, in fact, to be simply discrete interstellar clouds in the
remnant's vicinity. Rather than invoking a low-density, progenitor wind-driven
cavity, we find the remnant to be located in a broad region some eight degrees
off the Galactic plane that is relatively free of interstellar clouds.

Specifically, we conclude that the remnant's gross morphology is due to the
supernova's blast wave encountering a large molecular cloud off to its west
giving rise to NGC~6960, smaller discrete clouds off to the east and northeast
resulting in NGC~6992, NGC~6995, and IC~1340, and relatively dense clouds to
the northwest resulting in Pickering's Triangle and neighboring nebulae.
Although X-ray abundance data suggests the Cygnus Loop is the remnant of a
high-mass progenitor star, there is no need to propose a wind-driven cavity to
explain the remnant's low density interior but relatively dense emission limbs.

\begin{figure*}[t] 
\centering 
\includegraphics[width=0.90\textwidth]{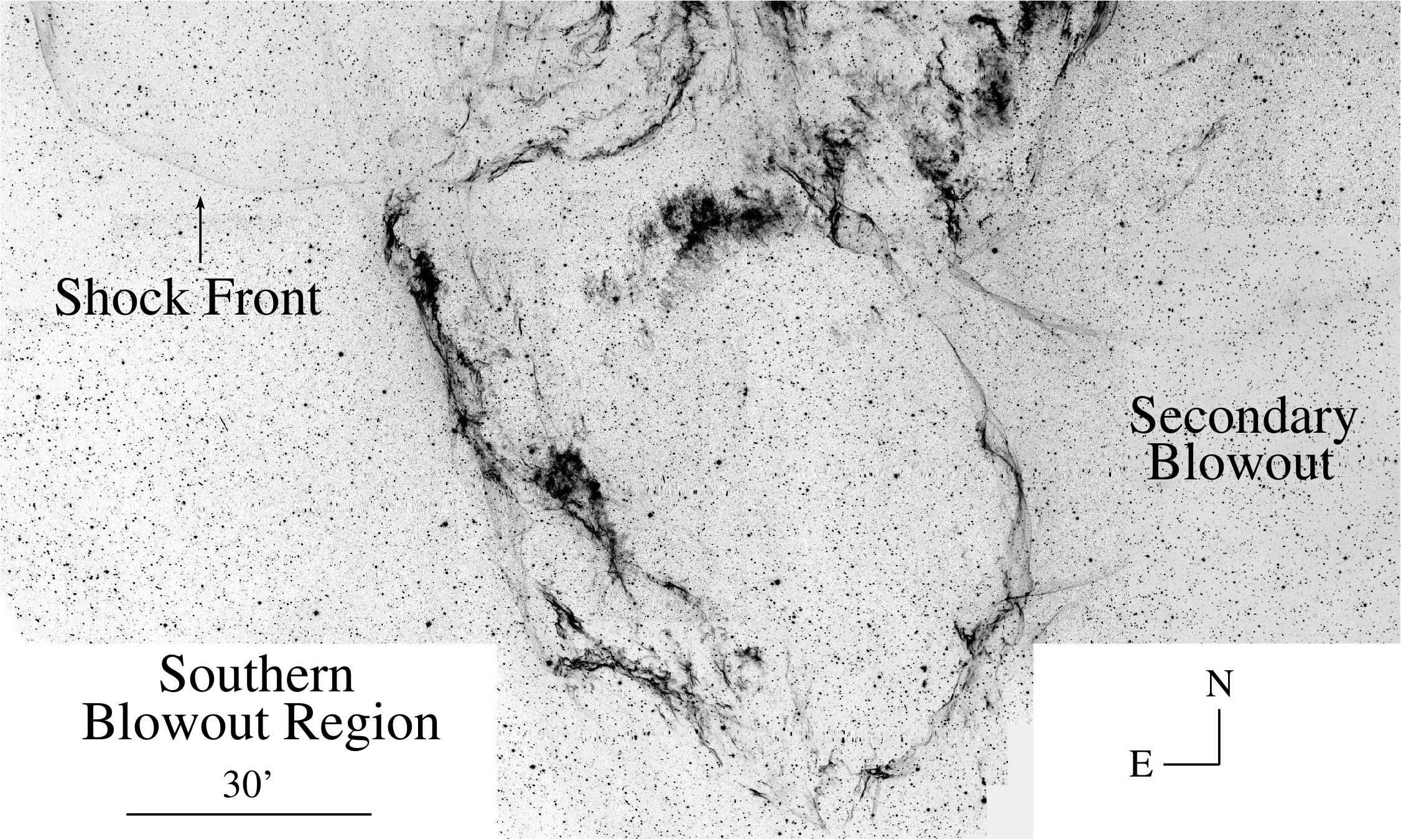} \\
\caption{Deep H$\alpha$ image of the Cygnus Loop's southern blowout region
showing a secondary blowout structure off to the west.  } 
\label{blowout_image}
\end{figure*}

\subsection{The Nature of the Remnant's Southern Blowout}

Finally, we address the nature of the southern blowout region which marks a
significant departure from the remnant's otherwise fairly spherical morphology
(see Fig.\ 1).  The blowout region's optical emission is better seen in Figure
9 which shows a mosaic of H$\alpha$ images taken with the 0.6 m Burrell Schmidt
telescope at Kitt Peak as part of a survey of the remnant's optical structure
(see \citealt{Pat2002} for details).  Although viewed most often as a
breakout of the remnant's expansion into a lower density region to the south
\citep{Levenson1998,Aschen1999}, or possibly the wake left behind the high-mass
progenitor as it moved northward \citep{Meyer2015}, there have been recent
suggestions that it constitutes a completely separate remnant. 

Based on radio continuum and polarization mapping at 2695 MHz that showed a
significantly larger percentage of polarization in the blowout region compared
to the remnant's northern part, \citet{Uyan2002} argued that the blowout
actually represented a separate SNR interacting with the Cygnus Loop. In
support of this scenario, they cited  the presence of a possible neutron star
near the centre of the blowout \citep{Miyata2001}, and differences between the
northern and southern parts of the Cygnus Loop in terms of optical and X-ray
emissions.  Although they noted difficulties explaining the lack of strong
X-ray emission in the overlapping and presumed interacting region, they
concluded that it was not just two SNRs seen in superposition, but actually in
physical contact.

Follow-up radio polarization studies by \citet{Sun2006} and
\citet{West2016} confirmed large differences in polarization characteristics
between the northern and southern portions of the Cygnus Loop, supporting the
notion of two separate remnants. However, an X-ray study by \citet{Uchida2009}
found no evidence in X-rays that the blowout region constitues a separate remnant. 

Likewise, there  there is no indication from optical images that the lower
portion of the Cygnus Loop's emission is separate or distinct from the rest of
the remnant (see Fig.\ 9).  The remnant's southern filamentary features connect
to filaments in the blowout region, with the Cygnus Loop's highly curved
southeastern shock front showing an abrupt cutoff at the intersection with the
blowout region. 

Moreover, the blowout region has its longer N-S axis and its southernmost tip
in rough alignment with the centre line of the Cygnus Loop's main shell. This
orientation is especially clear in X-ray maps of the remnant
\citep{Levenson1997,Aschen1999} and is not unlike that expected for a blowout of the
remnant's shock front encountering a lower density interstellar region.

Also as can be seen in Figure 9, there is a large secondary blowout off the western
side of the main blowout. The location of this secondary blowout roughly
coincides with a break in the southern extent of the western molecular cloud
(see Fig.\ 7) suggesting the blowout is at the same distance as the Cygnus Loop.  

Consequently, we find no morphological evidence to support a
second SNR scenario.  This conclusion is strengthened by the lack of X-ray
emission at a presumed interface region where two separate SNRs would be
interacting. Thus we conclude, as did \citet{Ku1984}, that the southern blowout
is simply the result of the remnant's shock encountering a low density region
(n$_{0}$ $< 0.1$ cm$^{-3}$) more than nine degrees off the Galactic plane
relatively absent of large interstellar clouds.

\section{Conclusions}

While the Cygnus Loop is among the best studied Galactic supernova remnants,
its distance has long been uncertain.  Here we present the first detections of
high-velocity interstellar absorptions of \ion{Na}{1} and \ion{Ca}{2} seen in
the spectra of stars with projected locations within the remnant's boundary.
These detections, along with the recent discovery of a  B7 star, BD+31~4224,
lying either inside and/or interacting directly with the remnant's expanding
blast wave along the Cygnus Loop's northwestern limb \citep{Fesen2018}, are
used to set the most robust distance estimate to the remnant. Our findings are
sumarized below. 

1) Moderate-dispersion spectra of three stars with projected locations toward
the remnant reveal the first detections of \ion{Na}{1} 5890,5896 \AA \ and
\ion{Ca}{2} 3934 \AA \ absorption features associated with the remnant's
expanding shell. Measured LSR velocities range from $-160$ to +240 km s$^{-1}$.

2) We estimate the centre of the Cygnus Loop lies at a distance of $735 \pm 25$
pc based on the detection of high-velocity \ion{Na}{1} absorptions in these
three background stars, combined with the distance to the B7 star interacting
with the remnant's shock.  We further find that the remnant is likely
aspherical in shape, with its eastern limb nearer to us than its northwestern
limb, with a diameter $\simeq$ 37 pc.

3) Using our new estimated distance for the Cygnus Loop, we calculate several
fundamental physical properties of the remnant which are in good agreement with
inferred values reported from numerous previous studies (Table 3).

4) From inspection of multi-wavelength composite images, we find the Cygnus
Loop's morphology to be the result of its location far off the Galactic plane
in a relatively low-density region in between local interstellar clouds.  The
interaction of the remnant's blast wave with these clouds explains the
locations of all its major optical and UV emission nebulae.  Consequently, we
propose that the Cygnus Loop is not a remnant inside a progenitor wind-driven,
low-density cavity but rather lies in an extended, low-density region in
between a dense molecular cloud to its west and northwest, with smaller clouds
to its east and northeast.

5) These composite images along with deep H$\alpha$ images show no evidence
supporting the notion that the Cygnus Loop's southern blowout region is a
separate SNR. Instead, the blowout is just what it appears to be: a blowout of
the Cygnus Loop's blast wave toward a low-density region largely absent of
interstellar clouds.

Finally, we note the star, TYC 2688-2556-1, identified by \citet{Boubert2017}
as a possible runaway former companion of the Cygnus Loop's progenitor at
estimated distances around 430 pc (Gaia DR2) is unlikely if our estimate of
$\simeq$735 pc is correct.  Likewise, the pulsar wind nebula candidate
discovered by \citet{Katsuda2012} in the remnant's southern blowout region, if
physically related to the Cygnus Loop, would require a transverse velocity in
excess of 1200 km s$^{-1}$, quite high but still lower than their estimate of
1850 km s$^{-1}$ assuming a distance of 540 pc and an age of 10,000 yr.

\acknowledgements

We thank the MDM Observatory staff for their excellent assistance and rapid instrument
re-configuration, and Ronald Downes and Debra Wallace for help in constructing an
H$\alpha$ survey of the Cygnus Loop.  This research was made possible in part
by funds from NASA's SpaceGrant, the Guarini School of Graduate and Advanced
Studies at Dartmouth, and is part of RAF's Archangel Research Program.  We have
made use of data from the European Space Agency (ESA) mission {\it Gaia}
(\url{https://www.cosmos.esa.int/gaia}), processed by the {\it Gaia} Data
Processing and Analysis Consortium (DPAC,
\url{https://www.cosmos.esa.int/web/gaia/dpac/consortium}). Funding for the
DPAC has been provided by national institutions, in particular the institutions
participating in the {\it Gaia} Multilateral Agreement.  This research has also
made use of the SIMBAD database, operated at CDS, Strasbourg, France.

\newpage

\end{document}